\documentclass[aps,prb,twocolumn,superscriptaddress,11pt,reprint]{revtex4-2}

\usepackage{float}
\usepackage{multirow}
\usepackage{array} 
\usepackage{graphicx}
\usepackage{bm}
\usepackage{color}
\usepackage{csquotes}
\usepackage{amssymb}
\usepackage{braket}
\usepackage{placeins}
\usepackage{subcaption}
\usepackage{xspace}
\usepackage[justification=raggedright,singlelinecheck=false]{caption}
\usepackage[colorlinks=true,backref=false, linktocpage=true,
citecolor=blue,urlcolor=blue,linkcolor=blue,pdfpagemode=UseOutlines]{hyperref}
\usepackage{cleveref}

\newcommand{\vs}{\texttt{vSineKAN}\xspace}
\newcommand{\rs}{\texttt{rSineKAN}\xspace}
\newcommand{\vm}{\texttt{vMLP}\xspace}
\newcommand{\rmlp}{\texttt{rMLP}\xspace}
\newcommand{\mlp}{\texttt{MLP}\xspace}
\newcommand{\lstm}{\texttt{RNN-LSTM}\xspace}
\newcommand{\rbm}{\texttt{RBM}\xspace}
\newcommand{\sk}{\texttt{SineKAN}\xspace}

\begin{document}

\newcommand*{\UA}{Department of Physics and Astronomy, University of Alabama, Tuscaloosa, 35487, Alabama, USA.}\affiliation{\UA} 
\newcommand*{\KU}{Department of Physics and Astronomy, University of Kansas, Lawrence, Kansas 66045, USA.}\affiliation{\KU} 
\newcommand*{\DU}{Department of Physics, University of Dhaka, P.O. Box 1000, Dhaka, Bangladesh.}\affiliation{\DU} 

\title{Probing Quantum Spin Systems with Kolmogorov-Arnold Neural Network Quantum States}

\author{Mahmud Ashraf Shamim}
\email{mashamim@crimson.ua.edu}
\author{Eric A. F. Reinhardt}
\email{eareinhardt@crimson.ua.edu}
\affiliation{\UA}
\author{Talal Ahmed Chowdhury}
\email{talal@ku.edu}
\affiliation{\KU}\affiliation{\DU}
\author{Sergei Gleyzer}\affiliation{\UA}
\author{Paulo T Araujo}\affiliation{\UA}

\begin{abstract}
\noindent Neural Quantum States (NQS) are a class of variational wave functions parametrized by neural networks (NNs) to study quantum many-body systems. In this work, we propose \sk, a NQS \textit{ansatz} based on Kolmogorov-Arnold Networks (KANs), to represent quantum mechanical wave functions as nested univariate functions. We show that \sk wavefunction with learnable sinusoidal activation functions can capture the ground state energies, fidelities and various correlation functions of the one dimensional Transverse-Field Ising model, Anisotropic Heisenberg model, and Antiferromagnetic $J_{1}-J_{2}$ model with different chain lengths. In our study of the $J_1-J_2$ model with $L=100$ sites, we find that the \sk model outperforms several previously explored neural quantum state \textit{ansätze}, including Restricted Boltzmann Machines (RBMs), Long Short-Term Memory models (LSTMs), and Multi-layer Perceptrons (MLP) \textit{a.k.a.} Feed Forward Neural Networks, when compared to the results obtained from the Density Matrix Renormalization Group (DMRG) algorithm. We find that \sk models can be trained to high precisions and accuracies with minimal computational costs.
\end{abstract}

\maketitle

\section{Introduction}\label{sec:intro}

Understanding the properties of strongly correlated many-body quantum systems is a major challenge in modern physics. This challenge arises from the complexity of many-body interactions and the exponential growth of the Hilbert space, making exact encoding of the ground state (GS) wave function nearly impossible. \textit{Variational} methods provide an effective approach to tackle this problem by employing parameterized trial wave functions and minimizing their energy expectation values, thereby enabling an efficient representation of quantum states \cite{Beccasf,Sorella}.

Recently, in a seminal work, Carleo and Troyer~\cite{Carleo2017} introduced a variational \textit{ansatz} in which a neural network (NN) parametrizes the many-body wave-function. This approach, known as the NN quantum state (NQS), leverages quantum Monte Carlo methods \cite{Sorella} and the expressive power of NNs \cite{Lu2017} to capture complex quantum correlations. This work shows that shallow networks, such as restricted Boltzmann machines (\rbm), can efficiently approximate the GSs of the quantum Ising and antiferromagnetic Heisenberg models, and are also able to capture volume-law entanglement entropy \cite{dong}, opening a new avenue for representing variational wave functions.

As a subsequent development, various NQS architectures, such as Recurrent Neural Networks (RNNs)~\cite{RNN1,RNN2,RNN3,RNN4}, Convolutional Neural Networks (CNNs)~\cite{CNN1, CNN2, CNN3, CNN4}, Autoregressive Models~\cite{ARN1,ARN2}, and Vision Transformers (ViTs)~\cite{ViT1,ViT2,ViT3,ViT4,ViT5,ViT6,ViT7}, have been proposed, each with varying degrees of effectiveness and expressibility, along with hybrid approaches that combine traditional variational wave functions and NNs~\cite{Nomura,Ferrari}. For a recent discussion and review of the expressibility of the NQS, please see Refs.~\cite{Yang_Soleimanifar, Lange_2024}.

NQSs are generally based on architectures involving feature extraction followed by Multi-Layer Perceptrons (MLPs) \cite{rumelhart1986learning}. These classic MLPs are feedforward neural networks with nonlinear activation functions.

\begin{figure*}[htbp]
    \centering
    \includegraphics[width=0.8\textwidth]{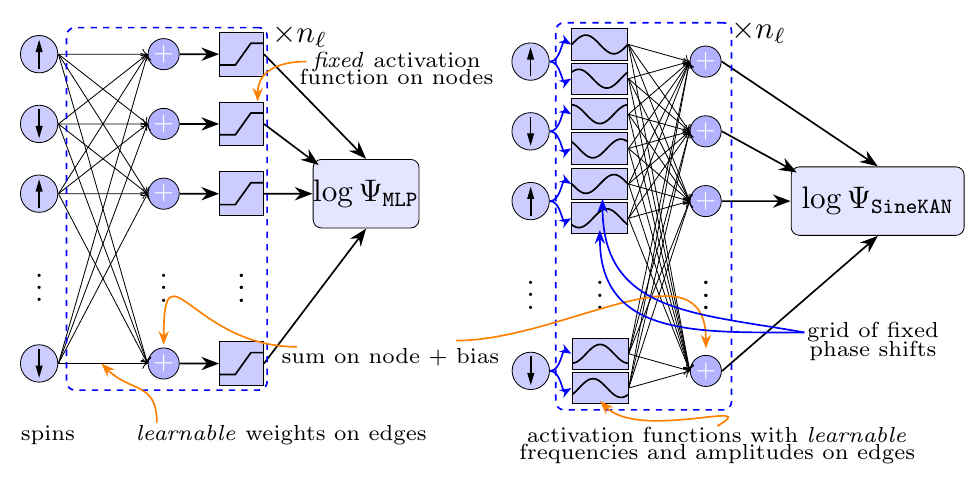}
\caption{\textbf{Left:} NQS based on an \mlp with $n_{\ell}$ layers. Spins are encoded as input, combined with learnable weights and biases, and passed through fixed activation functions to produce the wavefunction $\Psi_{\texttt{MLP}}$.  
\textbf{Right:} NQS based on \sk, where spins are passed through sinusoidal activations with learnable frequencies and amplitudes. A fixed grid of phase shifts is applied across $n_{\ell}$ layers, and outputs are combined to construct $\Psi_{\texttt{SineKAN}}$. In both cases, the learnable parameters are optimized using VMC (\autoref{sec:vmc}) to minimize the expectation value of the Hamiltonian $H$.}
    \label{fig:KAN}
\end{figure*}

In a recent work, Liu \textit{et al.} introduced Kolmogorov-Arnold Networks (KANs)~\cite{liu2024kan}, offering a promising alternative to traditional MLPs. These networks are motivated by the Kolmogorov-Arnold superposition theorem (KAST) that states that any continuous multivariate function can be represented by a two-layer model composed of univariate functions~\cite{Kolmogorov1956, Kolmogorov1957}. The original work explored the use of basis-splines as the learnable, on-edge function. However, a broad class of basis functions are possible, which has motivated the exploration of different KAN architectures beyond the original basis-spline \textit{ansatz}~\cite{liu2024kan}. Such work has included rational functions~\cite{rkan}, radial basis functions~\cite{rbfkan}, wavelet functions~\cite{wavkan}, Jacobi basis functions~\cite{fractionalkan}, Chebyshev polynomials~\cite{chebykan} and sinusoidal basis functions~\cite{sinekan}. Numerous follow-up works have shown the viability to various degrees of different KAN architectures across a broad range of tasks~\cite{comprehensiveKAN}.

In this paper, we explore a new class of NQS architecture that utilizes a KAN with sinusoidal activation functions, called the \sk \textit{ansatz} (see~\cref{fig:KAN}). \sk~\cite{sinekan} has already demonstrated superior performance in various tasks compared to other KAN architectures, such as image classification, in terms of numerical accuracy and computational efficiency. The primary motivation for selecting \sk as a variational wave function lies in its periodic and smoothly differentiable functional form, which makes it an effective approximator for a variety of functions. We benchmark the \sk \textit{ansatz} by computing the GS energies, fidelity and various correlation functions of the 1D transverse field Ising model (TFIM), anisotropic Heisenberg model (AHM)~\cite{Lieb}, and antiferromagnetic Heisenberg $J_{1}-J_{2}$ model. In each model for different chain lengths, we find excellent agreement in GS energies and correlation functions determined by Exact Diagonalization (ED) and the state-of-the-art Density Matrix Renormalization Group (DMRG) methods, and \sk which required only a modest number of variational parameters.

The paper is organized as follows. In section~\ref{sec:kan}, we introduce the Kolmogorov Arnold variational wave function and discuss different models we use for benchmarking our architectures in section~\ref{sec:bench}. The Hamiltonians we study are discussed in section~\ref{sec:ham}. In Section~\ref{sec:hyperparams} we discuss the training methods and the hyperparameters for different Hamiltonians. In section~\ref{sec:results}, we present the results of our simulations. Section~\ref{sec:msr} is dedicated to the Marshall Sign Rule (MSR) for \sk, and in section~\ref{sec:inference}, we discuss comparative runtime of our model with different models. Section~\ref{sec:discussion} provides a discussion of the findings, and section~\ref{sec:conclusion} concludes the paper with a summary and an outlook on future research directions.

\section{Kolmogorov-Arnold Variational Wavefunction}\label{sec:kan}
For a quantum mechanical system with $L$ sites, where each site is occupied by a single spin-\(\frac{1}{2}\), the dimension of the local Hilbert space is two, corresponding to the two possible spin states. In the computational basis, the spin configuration of the system is represented as $\sigma = \{ \sigma_1, \sigma_2, \cdots, \sigma_L \}$, where each $\sigma_i$ can take the values $\uparrow$ or $\downarrow$. A pure quantum state in this Hilbert space is a vector that can be expressed in terms of the basis states $\ket{\sigma}$ as
\begin{equation}
\Psi = \sum_{\sigma} \psi_{\sigma} \ket{\sigma}
\end{equation}
where $\psi_{\sigma}$ are the probability amplitudes associated  to $\ket{\sigma}$.

KAST states that any multivariate continuous functional mapping can be represented by sums of univariate functions of one variable~\cite{Kolmogorov1956, Kolmogorov1957}. As the wave functions are continuous, we can \textit{generalize} this to the problem of mapping spin lattices to an associated wave function as:
\begin{equation}
    \Psi \equiv \Psi_{\mathrm{KA}}  = \sum_{q = 1}^{2L+1} \Phi_{q}\left( \sum_{p = 1}^{L} \phi_{q,p}(\sigma_{p})\right)
\end{equation}
where  the inner function $\phi_{q,p}:\{0,1\}^L \rightarrow \mathbb{R}$ and the outer function $\Phi_{q}: \mathbb{R} \rightarrow \mathbb{R}$, and $\sigma_p$ are the spins are adjusted to have values of either 0 or 1 \cite{Kolmogorov1956,Kolmogorov1957}. 

Although KAST guarantees the existence of the inner functions $\phi$ and outer function $\Phi$, it does not specify their explicit form or provide a method for determining them. The KAN model addresses this limitation by representing each univariate function with a tunable parametrization defined by a fixed number $\beta$ of learnable parameters per function. For input dimension $L$, there are $L(2L+1)$ inner functions and $2L+1$ outer functions, resulting in a total of $(2L+1)(L+1)$ univariate function instances \cite{liu2024kan}. Consequently, the total number of trainable parameters becomes $\beta(2L^2 + 3L + 1)$, transforming the original function space from $\mathbb{R}^{2^L}$ to a much more compact and scalable space $ \mathbb{R}^{\beta(2L^2 + 3L + 1)}$. Assuming there exists a family of univariate functions that can achieve a good approximation with a relatively small number of parameters $\beta \ll \mathcal{O}(L^2)$, then the total parameter space is reduced from $\mathcal{O}(2^L)$ to $\mathcal{O}(L^2)$. With this approximation, for a set of parameters $\boldsymbol{\theta} = (\theta_1, \theta_2, \cdots, \theta_{\ell})$ with $\ell = \beta(2L^2+3L+1)$, the exact KA wavefunction $\Psi_{\mathrm{KA}}$ can be approximated by a KAN representation as 

\begin{equation}
\Psi_{\mathrm{KA}}(\sigma) \approx \Psi_{\mathrm{KAN}}(\sigma;\boldsymbol{\theta}).    
\end{equation}

This KAN representation of the many-body wave function can effectively be treated as the variational wavefunction with optimizable parameters $\boldsymbol{\theta}$ and is referred to as KA neural network quantum states \textit{ansatz}. Given a Hamiltonian $H$ that describes local interactions among the spins, and $\Psi_{\texttt{KAN}}\equiv\Psi_{\mathrm{KAN}}(\sigma;\boldsymbol{\theta})$ as the variational wavefunction, its ground state energy can computed by minimizing
\begin{equation}
E_{\theta} = \frac{\bra{\Psi_{\texttt{KAN}}}H\ket{\Psi_{\texttt{KAN}}}}{\braket{\Psi_{\texttt{KAN}}|\Psi_{\texttt{KAN}}}} 
\end{equation}
using the standard Variational Monte Carlo (VMC) protocol (\autoref{sec:vmc}). Using this approach, the KAN parameters (\boldsymbol{$\theta$}) can be approximated even when sampling only a relatively small portion of the total $2^L$ Hilbert space. 

In our work, we use learnable \textit{sinusoid} activation functions both for the inner and outer functions based on the multi-layered \sk architecture \cite{sinekan}. \sk was shown in \cite{sinekan} to exhibit properties useful for the task of representing the wavefunctions. One such property is its ability to generalize and project periodic patterns across spaces of a large domain that it hasn't been fitted to (something highly likely to occur in probing a Hilbert space of dimension $2^L$). Additionally, it is both continuous and differentiable, which could lead to improved stability in modeling continuous quantum wave functions. We show later in this work that this \textit{ansatz} is empirically effective compared to other explicitly proven approximators.

The functional form of each \sk layer can be expressed as follows:

\begin{equation}
    y_m = \sum_{l,n}^{L,N} A_{mnl} \sin\left(\omega_n x_l + \frac{\pi n}{N} + \frac{\pi l}{L} + \delta_{nl}\right) + b_m
\end{equation}

where $l$ is in the input index, $n$ is the grid index, $m$ is the output index, $A_{lmn}$ are amplitudes, $\omega_n$ are frequencies, $N$ is the grid size, $l$ is the input size, and $\delta_{ln}$ is a small positive random perturbation along the input and grid dimensions. The frequencies, amplitudes and biases are real-valued and serve as the learnable parameters of the model (see~\cref{fig:KAN}).

The 1D spin lattice arranged into a chain with a periodic boundary condition (PBC) is invariant under translation and reflection about any point in the chain. Using this \textit{ansatz}, it's computationally expensive to impose translational invariance, however, invariance under a single reflection about the center can be easily achieved. To preserve reflection symmetry, the \textit{ansatz} becomes:

\begin{equation}
    y = S(\vec{\sigma}) + S(\mathcal{R}(\vec{\sigma}))
\end{equation}

where $\mathcal{R}$ is a reflection operator along the 1D input dimension, $S$ is the same multi-layer \sk model, and $\vec{\sigma}$ are the set of input spins with values $+1$ or $-1$. By imposing this symmetry, we can drastically reduce the total parameter space over which we need to search. To impose such a symmetry for translational invariance would scale the model size linearly with the length of the lattice. We will refer to a regular multilayer \sk without reflection as \vs and the version with reflection symmetry as \rs.

\section{Benchmark NQS Models}\label{sec:bench}
We evaluate the effectiveness of the \sk \textit{ansätze} by benchmarking against three widely used NQS baselines: the \rbm, \lstm, and \mlp, as described in \cref{sec:j1j1-params}.

As the original NQS proposed by Carleo and Troyer, \rbm has served as a foundational benchmark in variational studies of quantum many-body systems \cite{Carleo2017, smolensky1986harmony}. Its main advantage is that it's high computational efficiency for small to medium-sized systems, which allows it to scale to larger systems—up to around 100 spins—with reasonable training time, especially compared to more complex architectures like \lstm. The computational complexity of \rbm is more comparable to $\mathcal{O}(e^N)$, so they remain highly efficient as long as the lattice size is relatively small. Structurally, the \rbm consists of a visible layer representing spin configurations and a hidden layer of binary units, connected through trainable weights without intra-layer couplings (\cite{smolensky1986harmony, hinton2002training}). Due to this bipartite structure, \rbm is particularly effective in NQS applications for unfrustrated or weakly frustrated systems.

The application of RNNs to represent GS wavefunctions in quantum spin systems is well established, with several studies demonstrating their effectiveness~\cite{RNN1, RNN2}. The motivation behind the choice of RNN architectures is that they generally encode some information about long-term and short-term \enquote{memory} across the length of inputs in a sequence. The idea in this case is that the memory of the models could capture short- and/or long-range dependencies across the spin lattice. Here we use a long-short-term memory (\lstm) as compared to alternative GRU or simple RNN used in \cite{RNN1, RNN2} due to previous literature findings that \lstm generally outperform the other RNN models on large datasets and sequences where very-long-term memory is not necessary \cite{rnn_compare_1, rnn_compare_2}.

Finally, we compare against a \mlp \cite{rumelhart1986learning}. This architecture is nearly ubiquitous in modern machine learning. It's commonly used following some sort of feature extractor as in the case of CNN, RNN, and Transformers. We will benchmark against two versions of MLP, one we will call \vm which is a simple \mlp with ReLU activations in hidden layers and one called \rmlp which, similarly to \rs, includes a second forward pass of the reflected lattice. This is a natural comparison for the \sk to show the advantage provided by on-edge learnable activations compared to learnable on-edge weights and fixed activations on node.

\mlp (\cite{rumelhart1986learning}) is compared pictorially in \cref{fig:KAN}. The main differences are the fixed learning rate (typically ReLU \cite{relu}) and that all learnable weights and biases are applied on the edge of the graph before passing through the activation function.

\section{Model Hamiltonians}\label{sec:ham}
Our study focuses on evaluating how well the \sk captures the GS properties of the three different 1D quantum spin models: (1) TFIM, (2) AHM \cite{Lieb}, and (3) antiferromagnetic $J_{1}-J_{2}$ model, all with PBC. Their Hamiltonians are given by
\begin{equation}
    H_{\mathrm{TFIM}} = -J \sum_{i} \hat{\sigma}^{z}_{i}\hat{\sigma}^{z}_{i+1} - h \sum_{i} \hat{\sigma}^{x}_{i}
\end{equation}

\begin{equation}
    \hat{H}_{\gamma} = \sum_{i} [(1+\gamma) \hat{S}^{z}_{i} \hat{S}^{z}_{i+1} + (1-\gamma) (\hat{S}^{x}_{i} \hat{S}^{x}_{i+1} + \hat{S}^{y}_{i} \hat{S}^{y}_{i+1})]
\end{equation}
and
\begin{equation}
    \hat{H}_{J_{1}-J_{2}} = J_{1} \sum_{i} \hat{\mathbf{S}}_{i}.\hat{\mathbf{S}}_{i+1} + J_{2} \sum_{i} \hat{\mathbf{S}}_{i}.\hat{\mathbf{S}}_{i+2}
\end{equation}
where $\hat{\mathbf{S}}_{i} = (\hat{S}_{i}^{x}, \hat{S}_{i}^{y}, \hat{S}_{i}^{z})$ are the spin-$\frac{1}{2}$ operators acting on site $i$. The anisotropy parameter $-1\leq \gamma \leq 1$, $J_{1} > 0$ and $J_{2} \geq 0$ are the nearest and next-nearest-neighbor antiferromagnetic couplings, respectively. These two models become identical when $J_{1} = 1, J_{2} = 0$ and $\gamma = 0$.

The TFIM describes a system of spin-$\frac{1}{2}$ particles arranged on a 1D lattice with interaction between nearest-neighbor spins and an external magnetic field. The coupling $J$ in $H_{\mathrm{TFIM}}$ represents the ferromagnetic Ising interaction, which favors adjacent spins in the $z$-direction. The strength of the external magnetic field is given by $h$ which acts along the $x$-axis, inducing quantum fluctuations that compete with the ferromagnetic interaction. The $\hat{\sigma}^{x}_{i}$ and $\hat{\sigma}^{z}_{i}$ are the Pauli spin operators at the site $i$. In the thermodynamic limit, this model exhibits a quantum phase transition (QPT) in a critical magnetic field $h_{c} = J$ separating a ferromagnetic phase ($h < h_{c}$) and a paramagnetic phase ($h > h_{c}$). At the critical point, the system undergoes a continuous phase transition belonging to the 2D classical Ising model class.

The Hamiltonian $H_{\gamma}$ defines a 1D quantum spin-$\frac{1}{2}$ chain with anisotropic couplings in different spin directions.  The model is an exactly solvable free‐fermion chain at $\gamma=-1$, reduces to the $SU(2)$‐invariant Heisenberg chain at $\gamma=0$, and in the limit $\gamma\to1$ becomes the classical antiferromagnetic Ising chain in the $z$-basis with two-fold degenerate N\'eel product-state ground states.  In the thermodynamic limit the system lies in a gapless Luttinger-liquid phase with algebraic spin correlations for $\gamma\in[-1,0]$, whereas for $\gamma>0$ it develops a finite spin gap and N\'eel order; these two regimes are separated by a quantum critical point at $\gamma=0$.

Antiferromagnetic $J_{1}-J_{2}$ is a paradigmatic model for studying frustrated magnetism with rich GS properties.  Its phase diagram is well understood through analytical and numerical methods \cite{Okamoto,White1}. For $J_{2} = 0$ the model becomes exactly solvable by the celebrated Bethe \textit{ansatz} solution \cite{bethe1931theorie}. With an increasing value of $J_{2}$, it undergoes a quantum phase transition from the gapless spin-fluid phase to a spontaneously dimerized gapped valence bond state (VBS) phase. The critical value of $J_{2}$ has been determined with high accuracy as $J_{2}^{c} = 0.241167  \pm 0.000005$ \cite{Eggert}. The model becomes solvable at the Majumdar-Ghosh (MG) point $J_{2} = 0.5$ with a two-fold degenerate VBS GS \cite{MG1}.

For the Hamiltonians $\hat{H}_\gamma$ and $\hat{H}_{J_1 - J_2}$, when either $J_1 = 0$ or $J_2 = 0$, the spin chains are \textit{unfrustrated} and \textit{bipartite}, allowing a natural division into two sublattices, $A$ and $B$. In such cases, the GS wavefunction exhibits a characteristic sign structure known as the \textit{Marshall Sign Rule} (MSR)~\cite{marshall1955,Luca}. Explicitly, the GS can be written as:
\begin{equation}
    \Psi(\sigma) = (-1)^{N_A(\sigma)} \, \tilde{\Psi}(\sigma),
\end{equation}
where $N_A(\sigma) = \sum_{i \in A} \sigma_i$, with $\sigma_i = 0$ or $1$ encoding spin-down and spin-up respectively, and $\tilde{\Psi}(\sigma) \ge 0$ is a real, non-negative amplitude in the Ising basis. This sign structure can be eliminated by a unitary transformation $\hat{R}$, defined by: $\hat{R} \ket{\sigma} = (-1)^{N_A(\sigma)} \ket{\sigma}$, which acts as a basis rotation that absorbs the alternating sign from the wavefunction into the Hamiltonian. The rotated Hamiltonian, $\hat{H}_{\mathrm{rot}} = \hat{R} \hat{H} \hat{R}^{\dagger}$, then has strictly non-positive off-diagonal elements in the Ising basis. This guarantees that the GS of $\hat{H}_{\mathrm{rot}}$ is sign-free (i.e., positive-definite), by the Perron–Frobenius theorem.

Although the MSR is exact only in the unfrustrated case, it remains a remarkably accurate approximation for the sign structure of the ground state wavefunction up to moderate frustration ($J_2 \lesssim 0.5$), as confirmed by exact diagonalization (ED) studies~\cite{Beccarbm}. In our work, we explicitly encode this sign structure into the Hamiltonians used for variational learning. Additionally, for both the AHM and $J_{1}-J_{2}$ model we carry out all simulations in the zero total-magnetization sector of the Hilbert space.

\section{Training Algorithms and Hyperparameters} \label{sec:hyperparams}

In this section we discuss the training algorithms and hyperparameters used to train \sk architectures for the transverse field Ising model (TFIM) (\cref{sec:tfim-params}), anisotropic Heisenberg model (AHM) (\cref{sec:ahm-params}) and the $J_1-J_2$ model (\cref{sec:j1j1-params}). We consider the chain length $L = 20$ for TFIM, AHM, and for the $J_1-J_2$ model, we study the chain lengths $L =20, 32, 64$ and $100$.

\subsection{TFIM}\label{sec:tfim-params}
For all the computations of the TFIM, the spins are sampled with the \texttt{MetropolisLocal} algorithm. The sampler is configured with 1024 parallel Markov chains, and the number of samples and the chunk size are also set to 1024, ensuring that sampling is fully parallelized with no internal batching. A fixed random seed is used to ensure reproducibility. Both models used three \sk layers with a grid size of 8 and hidden layer dimensions \texttt{(64, 64)}. The wave function is optimized using the Adam optimizer, with a linearly decaying learning rate schedule defined via \texttt{optax}. Specifically, the learning rate starts at $10^{-4}$ for the first 5{,}000 training epochs and decays linearly to $10^{-6}$ over the remaining 5{,}000 epochs. This schedule allows for rapid initial convergence while ensuring stable updates in the later stages of optimization. For the zero–transverse‐field Ising point, we instead use the learning rate of $10^{-2}$ to avoid strong local minima and for 100 iterations since the ground state rapidly converges for this point.

\section{AHM}\label{sec:ahm-params}
\begin{figure}[ht]
    \centering
    \includegraphics[width=\columnwidth]{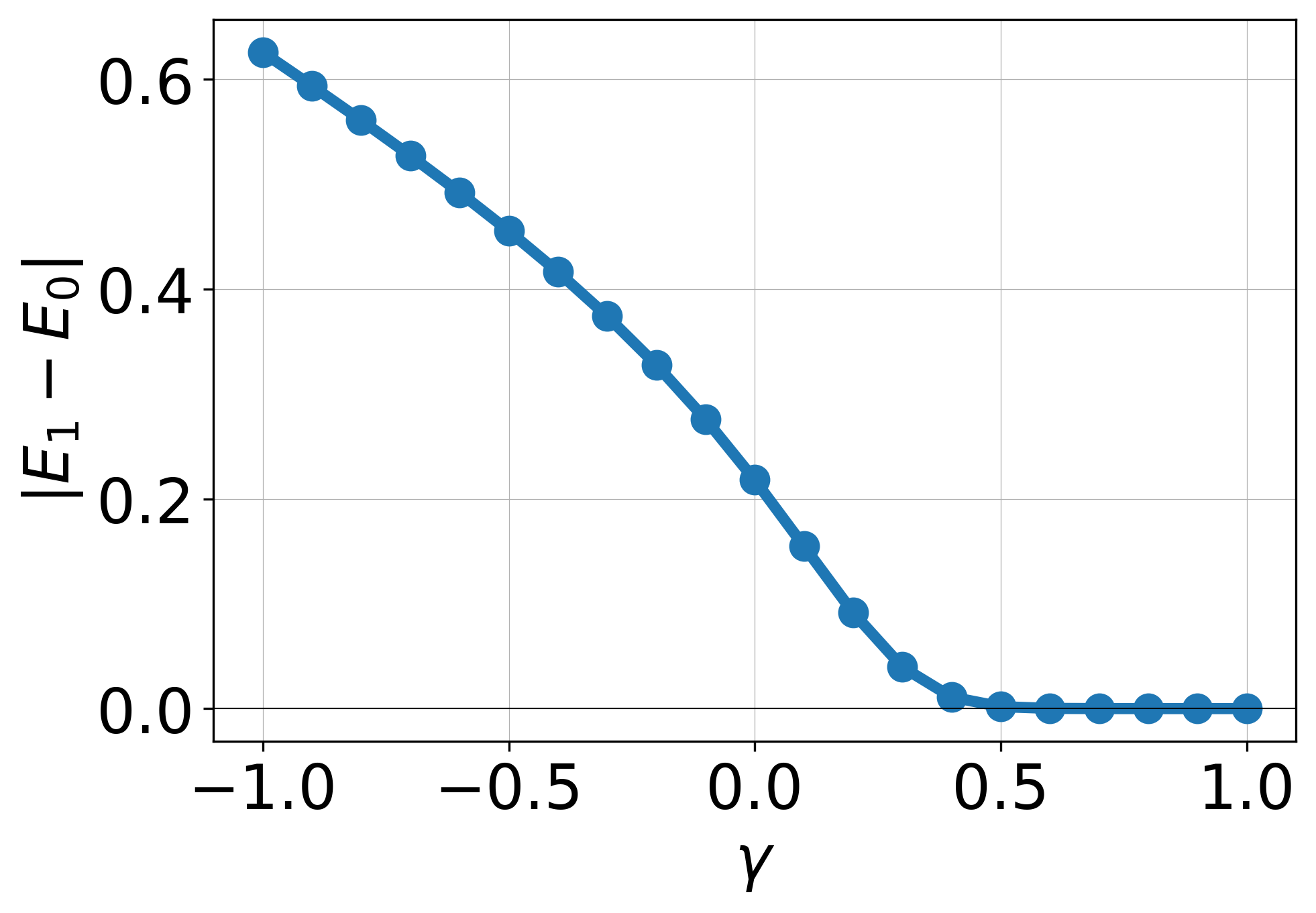}  
    \caption{Energy gap ($|E_{0}-E_{1}|$) between the GS and the $1^{\mathrm{st}}$ excited state for the $L=20$ anisotropic Heisenberg chain.}
    \label{fig:AHM gap}
\end{figure}

All sampling, ansatz, and optimizer settings are identical to those in Sec.~\ref{sec:tfim-params}. 
However, in the gapped phase as \(\gamma \to 1\), the spectral gap between the GS and the \(1^{\mathrm{st}}\) excited state decreases continuously and becomes vanishingly small by \(\gamma\approx0.6\) as shown in \cref{fig:AHM gap}.  Consequently, \sk with \texttt{MetropolisLocal} sampling becomes inefficient and gets trapped in a local minimum, leading to incorrect GS energies. We overcome this difficulty by adopting a \textit{Zeeman bias} \cite{Sandvik1999}. For $0 < \gamma < 0.9$, we augment
$H(h) = H_{\gamma} - h \sum_{i}\hat{S}^{z}_{i}$
while for $\gamma \ge 0.9$ (the Ising limit) we instead use $H(h) = H_{\gamma} - h \sum_{i}\hat{S}^{x}_{i}$.
This bias lifts the Néel degeneracy by an energy $\propto Lh$ and yields a unique nondegenerate GS.

The quench proceeds in two phases. First, we set the initial pinning strength to \(h_{\mathrm{init}} = \gamma + 0.2\). We then linearly ramp this field to zero in 15 equal steps: at each annealing stage \(n=0,\cdots,14\) we set $h_{n} = h_{\mathrm{init}}\Bigl(1-\frac{n+1}{15}\Bigr)$ and perform 333 VMC iterations under the Hamiltonian $H(h_{n})$, for a total of 4995 iterations. Once the field reaches $h=0$, we run an additional 5005 iterations on the bare Hamiltonian $H_{\gamma}$ to finalize the wavefunction in the true zero-field GS. For $\gamma$ between $-1$ to $0$ no pinning or bias was applied.

\subsection{$J_1-J_2$ Model}\label{sec:j1j1-params}
All \sk architectires are trained using the Adam optimizer \cite{Adam}. They are also trained with the same schedule of 30K epochs at a fixed learning rate followed by a decay down to 0.2 times the initial learning rate over 4K epochs. The learning rates for each model were found empirically by checking learning rates from $1\times 10^{-2}$ to $1\times 10^{-6}$ and the associated performance in evaluating GS energy values compared to DMRG. For \sk the learning rate was found to be $1\times 10^{-3}$ for values of $J_{2} < 0.3$ and $1\times 10^{-4}$ for values of $J_{2} \geq 0.3$, for \mlp it was found to be $1\times 10^{-3}$ for $J_2 < 0.4$ and $1\times 10^{-4}$ for $J_2\geq0.4$, for \lstm it was found to be $1\times 10^{-3}$, and for \rbm it was found to be $1\times 10^{-5}$.

The models are fitted using the NetKet \cite{vicentini2022netket} library with 1024 samples. For the \rbm and \sk models, the \texttt{local Metropolis-Hastings} algorithm was used while for the autoregressive direct sampler algorithm designed for use with the NetKet \lstm model was used.

The \vs and \rs models were constructed from three \sk layers with a grid size of 8 and output dimensions of 64, 64, and 1. The \vm and \rmlp models were constructed from three linear layers with output dimensions of 256, 256, and 1 with ReLU functions following the first two hidden layers \cite{relu}.

The \lstm model was constructed with three hidden layers each with a hidden dimension of 64. The \rbm was constructed with an alpha value (number of connections per feature) of 128. A table of model parameter numbers for lattice of size $L=100$ can be found in \cref{tab:params}.

\begin{table}[]
    \centering
    \begin{tabular}{|>{\centering\arraybackslash}p{4cm}|>{\centering\arraybackslash}p{4cm}|}
        \hline
         \texttt{Model} & \texttt{Parameters} \\
         \hline
         \rbm & 1,292,900 \\
         \sk & 86,433 \\
         \mlp & 91,905\\
         \lstm & 83,240\\
         \hline
    \end{tabular}
    \caption{Number of model parameters in \rbm, \sk, \mlp, \lstm with the system size $L=100$.}
    \label{tab:params}
\end{table}

For reasons explained in \cref{sec:discussion}, and as shown in \cref{tab:failcase}, a grid size of 7 is used rather than 8 for $L=64$.

\begin{table}
    \centering
    \begin{tabular}{|c|c|c|c|c|c|}
        \hline
        $L$ & \texttt{Grid Size} &  \texttt{Exact} & \texttt{Mean} & \texttt{Variance} & \texttt{Rel. Error} \\
        \hline
        64 & 8 & -24 & -16.962 & 5.317 & $2.932 \times 10^{-1}$ \\
        64 & 7 & -24 & -23.999 & $1.613 \times 10^{-5}$ & $3.698 \times 10^{-5}$ \\
        \hline
        100 & 10 & -37.5 & -34.671 & $4.954 \times 10^{-2}$ & 0.075 \\
        100 & 9 & -37.5 & -37.501 & $5.327 \times 10^{-7}$ & $1.492 \times 10^{-5}$\\
        \hline
    \end{tabular}
    \caption{Performance of two hidden layer \sk when hidden layer dimensions equal chain length ($L$) with a grid size equal to $\sqrt{L}$ versus $\sqrt{L}-1$.}
    \label{tab:failcase}
\end{table}

All models were trained on a single T4 GPU with 16GB of VRAM.

\begin{figure*}[htb]
    \centering
    \includegraphics[width=1.0\textwidth]{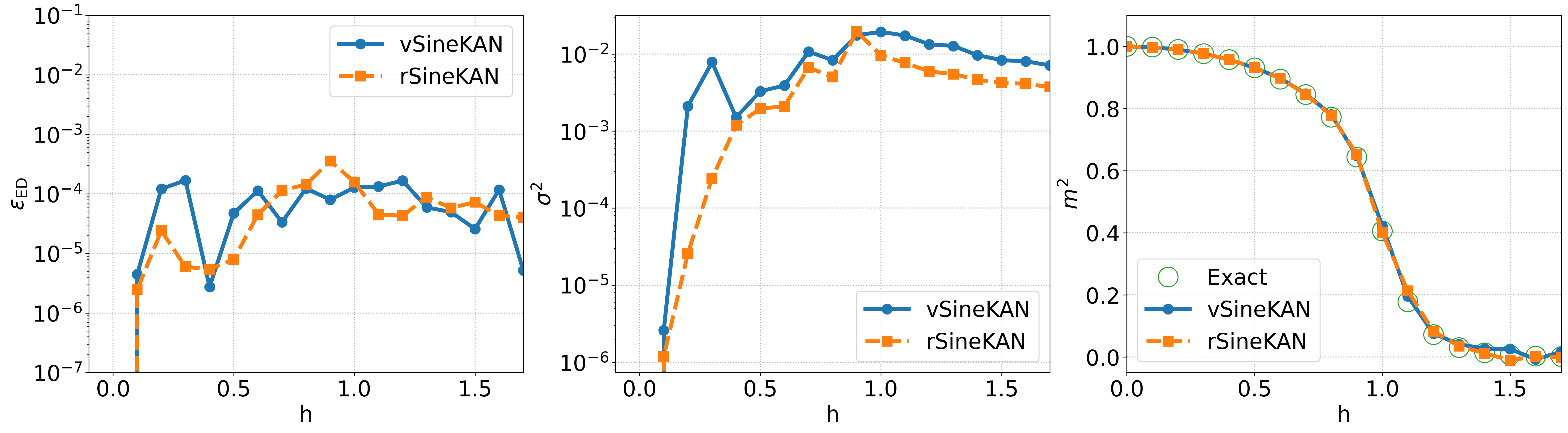}  
\caption{
        Comparison between the \vs and \rs \textit{ansatz} for the 1D TFIM with system size $L = 20$. 
        \textbf{(a)} Log-scale plot of the GS energy error $\epsilon_{\mathrm{ED}}$ as a function of transverse field $h$. 
        \textbf{(b)} Log-scale plot of energy variance $\sigma^2$, indicating the quality of the variational wave function. 
        \textbf{(c)} Spin-spin correlation function  $m^2$ as a function of $h$, with comparison to the ED. For $h=0$, the \rs and \vs solutions are exact and $\epsilon_{\mathrm{ED}}=0=\sigma^2$.
    }
\label{fig:TFIM}
\end{figure*}

\begin{figure}[htb]
    \centering
    \includegraphics[width=\columnwidth]{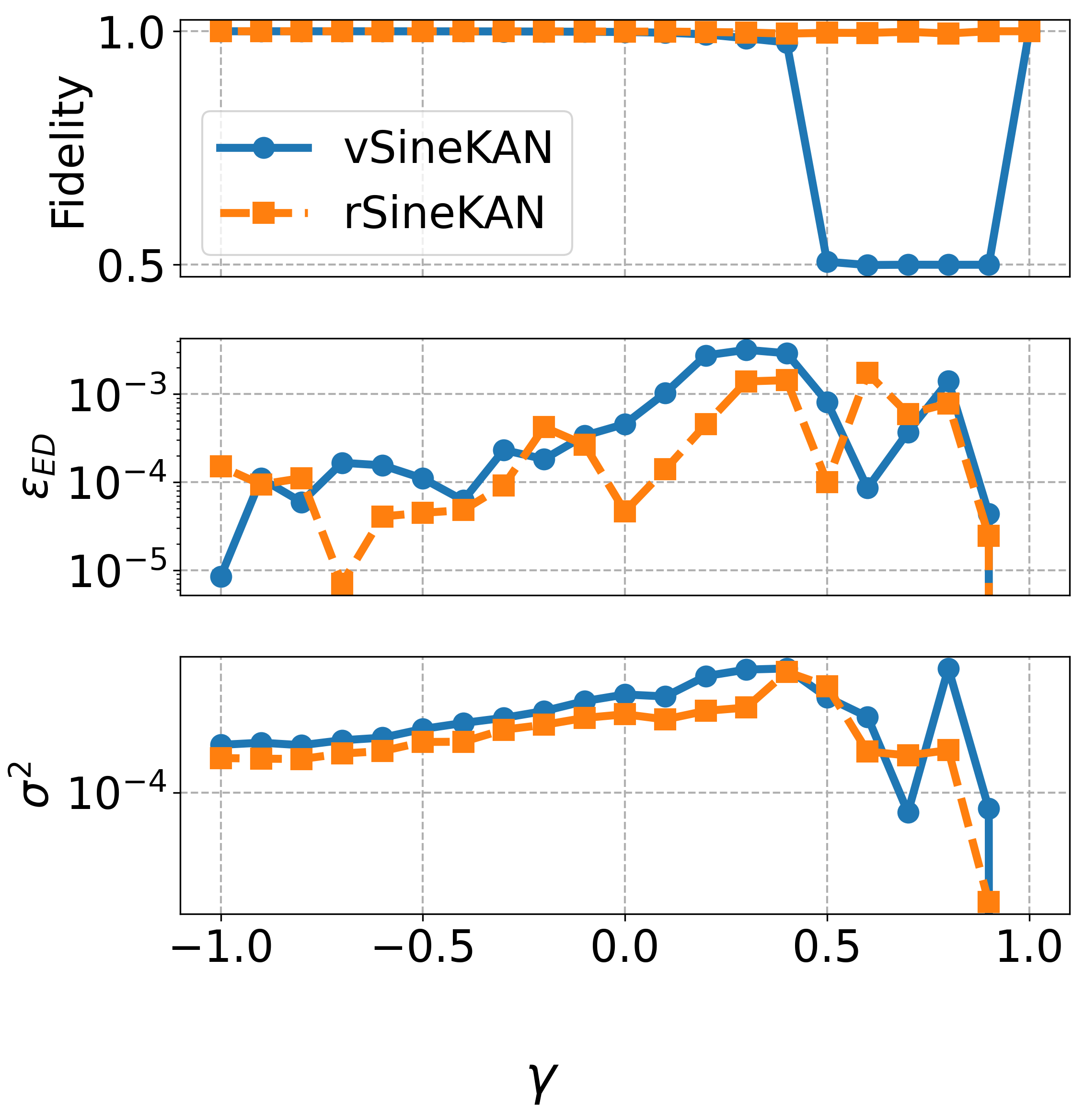}  
    \caption{
    Top: GS fidelity, Middle: Relative energy error, Bottom: Variance of the \vs and \rs ansatz compared to the exact GS energy as a function of $\gamma$ for the anisotropic Heisenberg chain of $L = 20$.
    }
    \label{fig:gamma}
\end{figure}

\section{Results} \label{sec:results}

\subsection{TFIM}
We begin by benchmarking the \vs and the \rs on the $L = 20$ transverse field Ising chain. For transverse field values $h \in [0.0, 1.6]$, we compute the GS energy, energy variance $\sigma^{2} \equiv \langle \hat{H}^{2} \rangle - \langle \hat{H} \rangle^{2}$, and the spin-spin correlation function defined as $m^{2}(h) = \frac{1}{L} \sum_{i} \langle \hat{\sigma}^{z}_{i} \hat{\sigma}^{z}_{i+L/2} \rangle_{h}$ and compare the results with ED. The relative error in energy is given by 
\begin{equation}\label{eq:epsed}
    \epsilon_{\mathrm{ED}} = \frac{|E_{\mathrm{ED}} - E_{\mathrm{Model}}|}{|E_{\mathrm{ED}}|}
\end{equation}
In \cref{fig:TFIM}, we plot the energy error and variance, as well as the spin-spin correlation $m^2(h)$, as functions of the transverse field $h$.

As shown in \cref{fig:TFIM}, both \vs and \rs \textit{ansatz} performs extremely well, achieving high energy accuracy and wave function quality, as indicated by the lower values of $\epsilon_{\mathrm{ED}}$ and $\sigma^{2}$  across the entire range of $h$. The spin-spin correlation $m^2(h)$ closely follows the ED results for both \textit{ansatz}, with the symmetric model providing slightly better agreement near the critical point $h_c \approx 1$. 

\subsection{AHM}
In the second benchmark of our \textit{ansatz}, we compute the GS energy, variance and fidelity of the AHM $H_{\gamma}$ for $\gamma\in[-1,1]$ and compare the GS energy with ED results for a chain of length $L = 20$. In \cref{fig:gamma}, we plot the relative energy error, variance and fidelity for different values of anisotropy.
For $-1\le \gamma\le 0$ (no Zeeman bias), both the \sk ansatz maintain $\varepsilon_{\mathrm{ED}}<5\times10^{-4}$. 

For $0 < \gamma \leq 1$, where a Zeeman bias is applied,  the near-critical window $0.1 \le \gamma \le 0.4$ produces the largest deviations: the \vs error peaks at \(\varepsilon_{\mathrm{ED}}(0.3)\approx3.2\times10^{-3}\) while the \rs remains below $\varepsilon_{\mathrm{ED}}(0.4)\approx1.4\times10^{-3}$.  As the bias reopens the gap for $\gamma \gtrsim 0.6$, both ansatz recover $\varepsilon_{\mathrm{ED}}<10^{-4}$, and by $\gamma\ge0.9$ both errors fall below $10^{-5}$, vanishing at $\gamma=1$.  

The fidelity $F$ between the exact GS and the \sk variational wavefunction is defined as:
\begin{equation}
F = \frac{|\langle \psi_{\text{ED}} | \psi_{\text{SineKAN}} \rangle|^2}{\langle \psi_{\text{ED}} | \psi_{\text{ED}} \rangle \langle \psi_{\text{SineKAN}} | \psi_{\text{SineKAN}} \rangle}. 
\end{equation}

As shown in the \cref{fig:gamma}, in the unbiased range $-1 \le \gamma \le 0$, both model achieve $F>0.99$ (e.g. $F_{r}(-1)=0.99997$, $F_{v}(-1)=0.99992$, and at $\gamma=0$, $F_{r}=0.99916$, $F_{v}=0.99833$).  In the biased region $0<\gamma<1$, however, the \vs partially fails to lift the Néel degeneracy despite quenching and its fidelity collapses to approximately $0.50$ for $0.5 \le\gamma\le 0.9$.  By contrast, the symmetric model maintains $F>0.99$ everywhere (e.g.\ $F_{r}(0.6)\approx0.996$, $F_{r}(0.8)\approx0.995$). At $\gamma = 1$ both fidelities recover to unity.  

As shown in the figure \cref{fig:gamma},  both ansatz maintain $\sigma^{2}<10^{-2}$, but the \rs keeps $\sigma^{2}<10^{-3}$ throughout, yielding an order‐of‐magnitude reduction in statistical fluctuations.

\subsection{Benchmark of $J_1-J_2$ Model Against ED}
In the third benchmark, we assess the accuracy of \sk in describing the GS of the 1D $J_1-J_2$ model through a comprehensive set of numerical test for system size $L = 20$. We evaluate the GS energy, the fidelity between ED and \sk GS wavefunction, as well as compute isotropic spin--spin and dimer--dimer correlation functions, and structure factors, comparing the results directly against ED. These comparisons test the ability of the \sk to faithfully represent not only the GS energy but also nontrivial correlation properties of the wavefunction for the frustrated spin system. 

We begin by evaluating the GS energy and fidelity between variational wavefunctions corresponding to different values of $J_2 = 0.0$ to $0.6$, using both the \vs and \rs ansatz. For the special case at the MG point at $J_2=0.5$, there is a two-fold degeneracy of the GS with identical energy of $-7.5$. The fidelity at this point is then calculated as the sum of the fidelity with each of the two orthonormal GS eigenvectors.

\begin{figure}[htbp]
    \centering
    \includegraphics[width=\columnwidth]{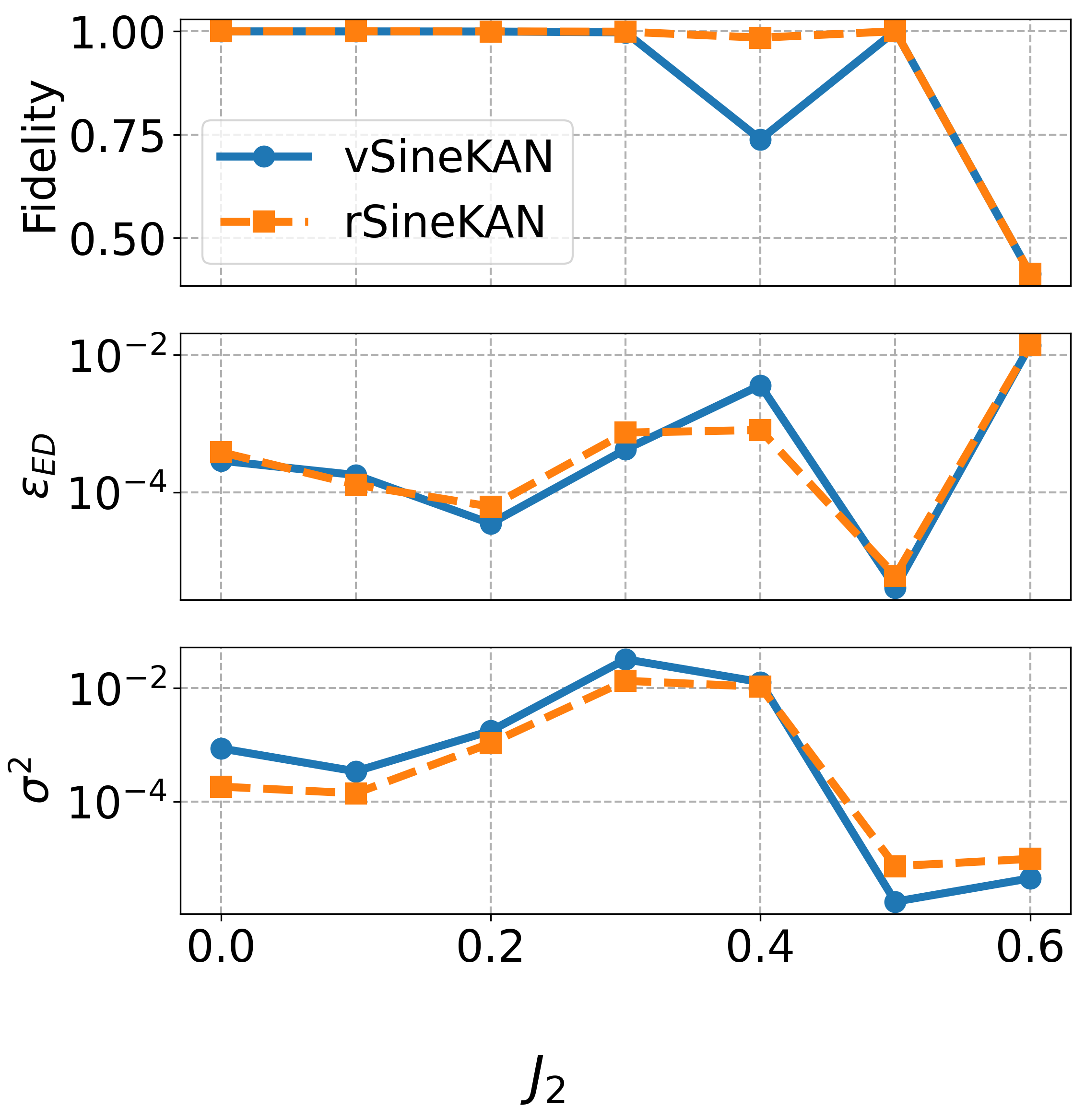}  
    \caption{
    Top: GS fidelity, Middle: Relative energy error, Bottom: Variance of the \vs and \rs ansatz compared to the exact GS energy as a function of $J_2$ for the Heisenberg $J_{1}-J_{2}$ chain of length $L = 20$.
    }
    \label{fig:Fidelity}
\end{figure}
As shown in \cref{fig:Fidelity}, the \rs model exhibits excellent fidelity with the exact GS across the unfrustrated and moderately frustrated regimes, specifically for  $0 \leq J_2 \leq 0.5$. In this range, the GS fidelity remains very close to unity, indicating that the \textit{ansatz} successfully captures the essential features of the GS wavefunction.

As shown in \cref{tab:j0p4fid}, for the case of $J_2=0.4$, the system is gapped, with a very small energy difference between the GS and the $1^{\mathrm{st}}$ excited state. The \vs model at this point shows a high fidelity ($\sim0.799$) with the GS and a fidelity of $\sim0.192$ with the $1^{\mathrm{st}}$ excited state with almost no contribution from higher excited states. Due to the small energy gap between the GS and $1^{\mathrm{st}}$ excited state, the GS energy of \vs of $-7.598$ represents an overlap between the two lowest energy states. \rs, however, is able reach a fidelity of $\sim1$ with the GS as seen in \cref{fig:Fidelity}. 

\begin{table}[ht]
    \centering
    \begin{tabular}{|>{\centering\arraybackslash}p{2.75cm}|>{\centering\arraybackslash}p{2.75cm}|>{\centering\arraybackslash}p{2.75cm}|}
         \hline
         \texttt{State} & \texttt{Fidelity} &  \texttt{ED Energy} \\
         \hline
         GS & 0.799 & -7.625\\
         $1^{\mathrm{st}}$ Excited & 0.192 & -7.575 \\
         $2^{\mathrm{nd}}$ Excited & 0.000 & -7.444\\
         $3^{\mathrm{rd}}$ Excited & 0.000 & -7.343\\
         \hline
    \end{tabular}
    \caption{Fidelity between \vs and ED GS and energy values associated with the first four excited states for $J_2=0.4$.}
    \label{tab:j0p4fid}
\end{table}

The \vs also performs well in the unfrustrated regime and retains high fidelity with the exact GS for $J_2 \leq 0.3$. In this region, the \textit{ansatz} is sufficiently expressive to capture the GS properties accurately. 

Interestingly, in the highly frustrated regime at $J_2 = 0.6$, both the \vs and \rs exhibit a sharp drop in fidelity, with values around $\sim 0.40\%$. This significant decline indicates that neither of our ansatz is able to faithfully represent the true GS in this regime. The poor performance can be attributed to the violation of the MSR, which is no longer satisfied $J_2 > 0.5$. In fact, the fidelity itself serves as a useful diagnostic tool to quantify the extent to which MSR is violated. A qualitative analysis of this case and its connection to MSR breakdown is presented in \autoref{sec:msr}. Since the \rs achieves higher fidelity compared to the \vs, we use the former throughout the rest of the section.

\begin{figure}[ht]
    \centering
    \includegraphics[width=\columnwidth]{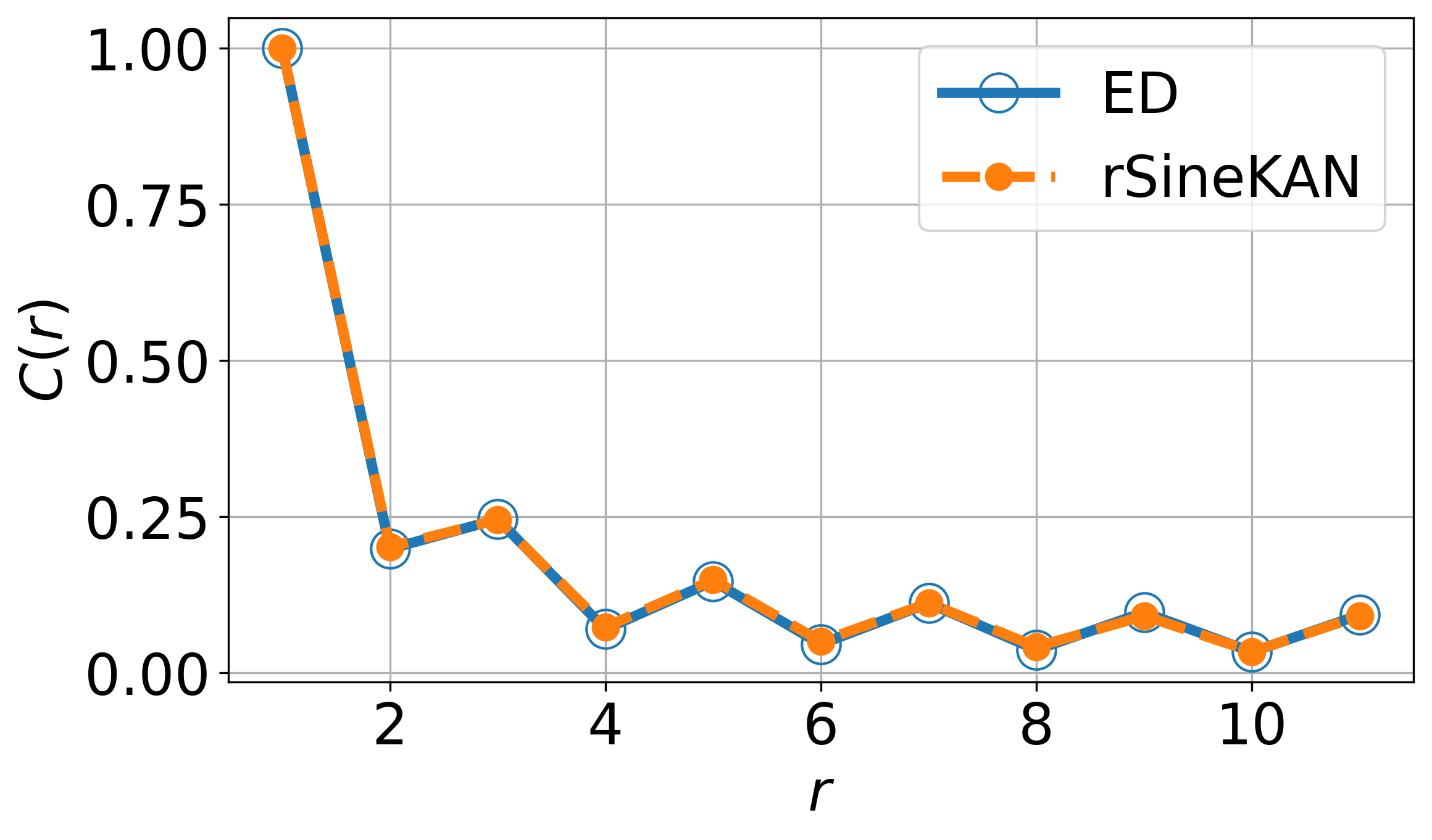}  
    \caption{Isotropic spin--spin correlation function $C(r)$ at $J_2 = 0.0$ for the chain of length $L = 20$, computed using ED and the \rs.}
    \label{fig:Isotropic spin}
\end{figure}

The spin-spin correlation functions are defined as
\begin{equation}
    C^{\gamma\gamma}(r) = \frac{1}{L} \sum_{\ell=1}^{L} \langle \hat{S}_{\ell}^{\gamma} \hat{S}_{\ell+r}^{\gamma} \rangle
\end{equation}
where, $\langle \cdots \rangle$ represents the expectation value taken over the variational wavefunction and $\gamma \equiv x, y$ or $z$. Following \cite{ViT1}, we compute the isotropic spin-spin correlations $C(r) = \frac{1}{3}(C^{xx}+C^{yy}+C^{zz})$, for $J_{2} = 0$, using \rs and compare its result against the ED for a system of $L = 20$. As shown in the \cref{fig:Isotropic spin}, the \rs reproduces ED results with high accuracy across all distances. 

\begin{figure}
    \centering
    \includegraphics[width=\columnwidth]{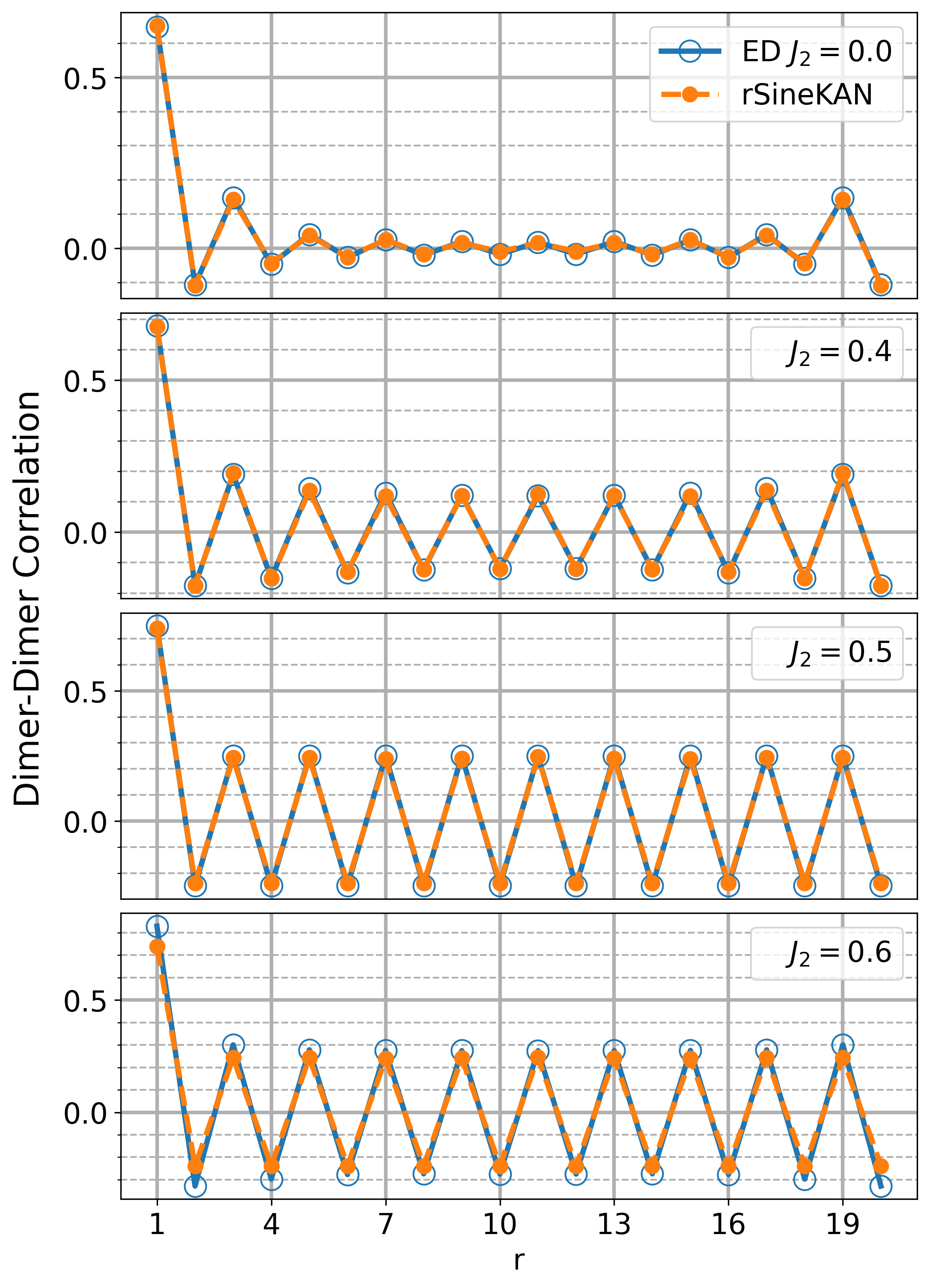}
    \caption{\rs results of the Dimer-Dimer correlations of $L=20$, $J_1-J_2$ model for (top to bottom): $J_{2}=0.0, 0.4, 0.5, 0.6$.}
    \label{fig:dimer-dimer}
\end{figure}

For $J_2 > J_{2}^{c}$, the system becomes gapped, with a dimerized two-fold degenerate GS that breaks translational invariance \cite{Sandvik}. The emergence of this dimer order can be probed using the connected dimer–dimer correlation function: \cite{ViT1}

\begin{equation}
    D(r) = \frac{1}{L} \sum_{\ell=1}^{L} \langle \hat{S}^{z}_{\ell}\hat{S}^{z}_{\ell+1} \hat{S}^{z}_{\ell+r}\hat{S}^{z}_{\ell+r+1}\rangle - [C^{zz}(r=1)]^{2}
\end{equation}

\noindent where, $C^{zz}(r=1) = \frac{1}{L}\sum_{\ell =1}^{L}\hat{S}^{z}_{\ell}\hat{S}^{z}_{\ell+1}$. For \rs and ED, the plot of $D(r)$ as a function $r$ for $J_{2} = 0, 0.4, 0.5$ and $0.6$ with system size $L=20$ are shown in \cref{fig:dimer-dimer}. There is a strong agreement in $D(r)$ between the ED solution and \rs for $J_2\le0.5$. The overall pattern is still preserved at $J_2=0.6$ but the correlation values start to diverge from ED.

\begin{figure}
    \centering
    \includegraphics[width=\columnwidth]{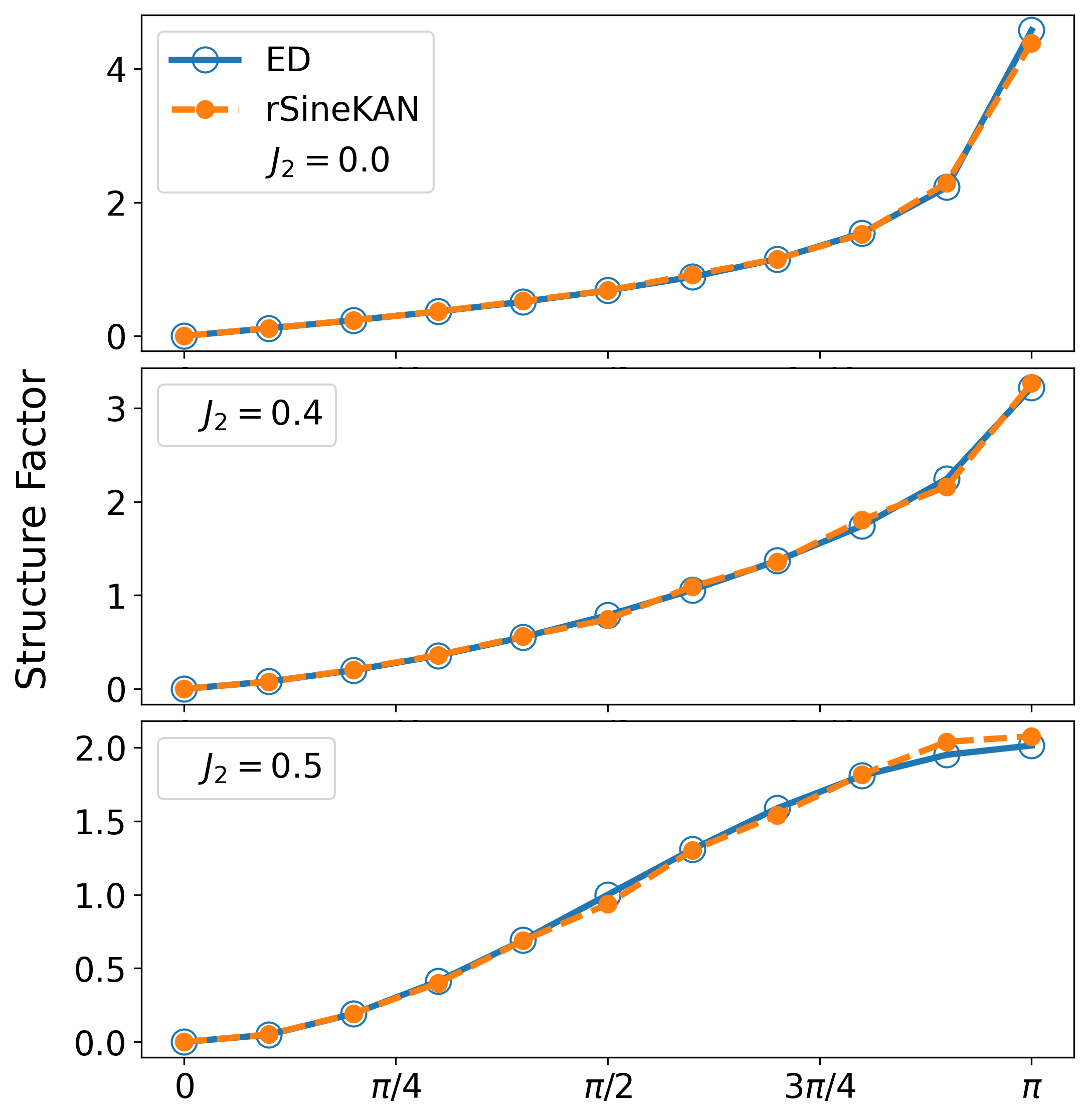}
    \caption{\rs results of the structure factor of $J_1-J_2$ model for (top to bottom): $J_{2}=0.0, 0.4, 0.5$.}
    \label{fig:structure-factor}
\end{figure}

The spin structure factor \cite{Beccasf} is defined as
\begin{equation}
    S(k) = \frac{1}{L}\sum_{i,j=1}^{L} e^{i(i-j)k} \langle 
    \hat{S}^{z}_{i}\hat{S}^{z}_{j}\rangle
\end{equation}
In the regime $J_2 \ll J_1$, strong quantum fluctuations prevent the system from developing true long-range order, which is reflected in a divergence of $S(k)$ at $k = \pi$. For the range $J_2^{c} < J_2 < 0.5$, spin fluctuations remain commensurate, meaning the peak of $S(k)$ stays at $k = \pi$. At the MG point, $J_2 = 0.5$, the system becomes dimerized and gapped, while still maintaining the peak of $S(k)$ at $k = \pi$. \cref{fig:structure-factor} shows the plots of $S(k)$ versus $k$ for $J_2 = 0, 0.4$ and $0.5$, obtained using both \rs and ED.
\subsection{Benchmarking $J_1-J_2$ Model for $L=32, 64$, \& $100$}\label{sec:j1j1-compare}

\begin{figure*}
    \centering
    \includegraphics[width=1.0\textwidth]{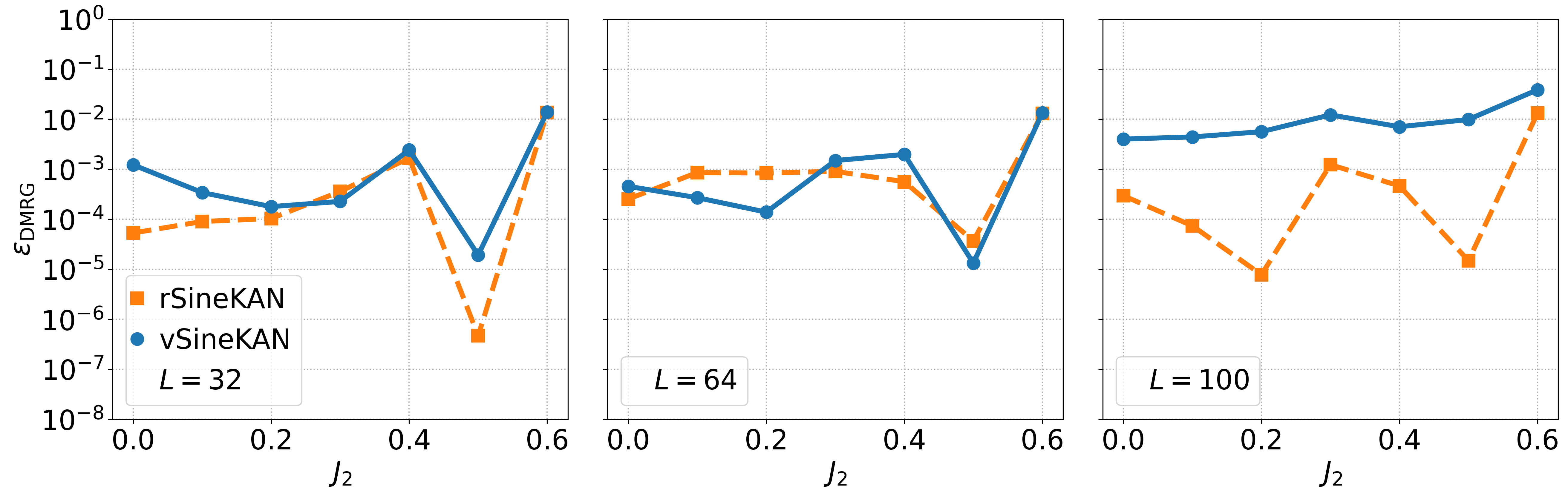}  
\caption{
Comparison of the relative error $\epsilon_{\mathrm{DMRG}}$ for the GS energy computed using the \vs (blue, solid line with circles) and the \rs (orange, dashed line with squares) as a function of $J_2$. The error $\epsilon_{\mathrm{DMRG}}$ is defined in (\ref{eq:1}). The three panels correspond to (left to right): $L=32$, $L=64$, and $L=100$. The reference energy is computed using DMRG with the maximum bond length $\chi = 200$.}
\label{fig:relerror}
\end{figure*}

\begin{figure*}
    \centering
    \includegraphics[width=1.0\textwidth]{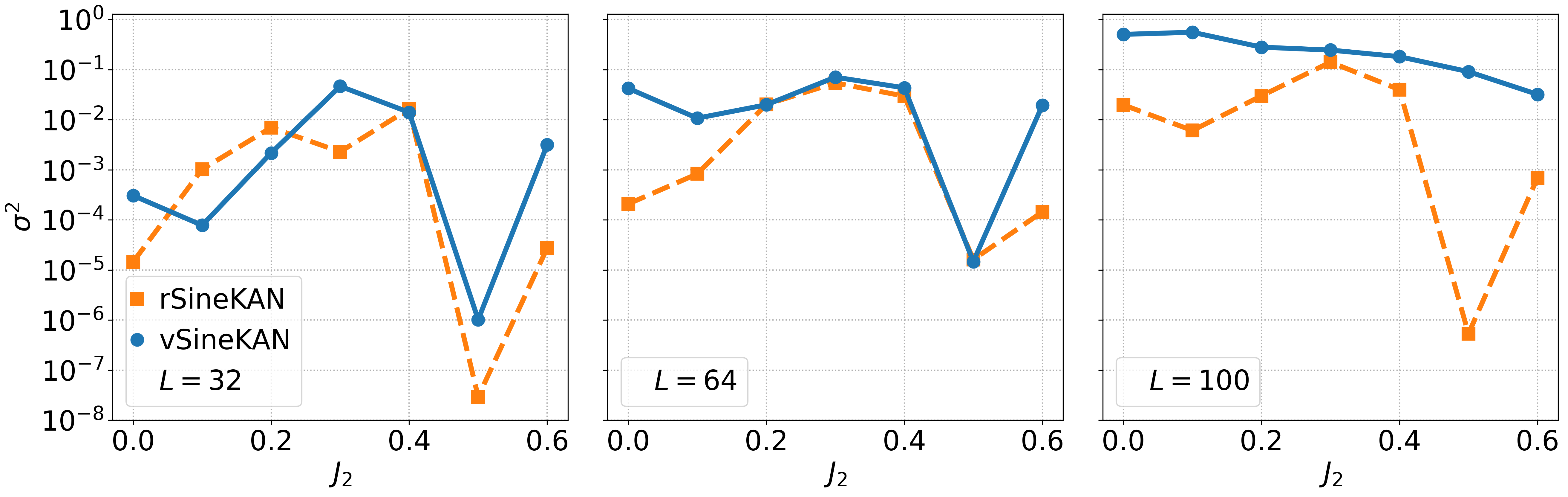}  
\caption{
Comparison of the variance for the GS energy computed using the \vs (blue, solid line with circles) and the \rs (orange, dashed line with squares) as a function of $J_2$. The three panels correspond to (left to right): $L=32$, $L=64$, and $L=100$.}
\label{fig:variance}
\end{figure*}

\begin{figure*}
    \centering
    \includegraphics[width=\linewidth]{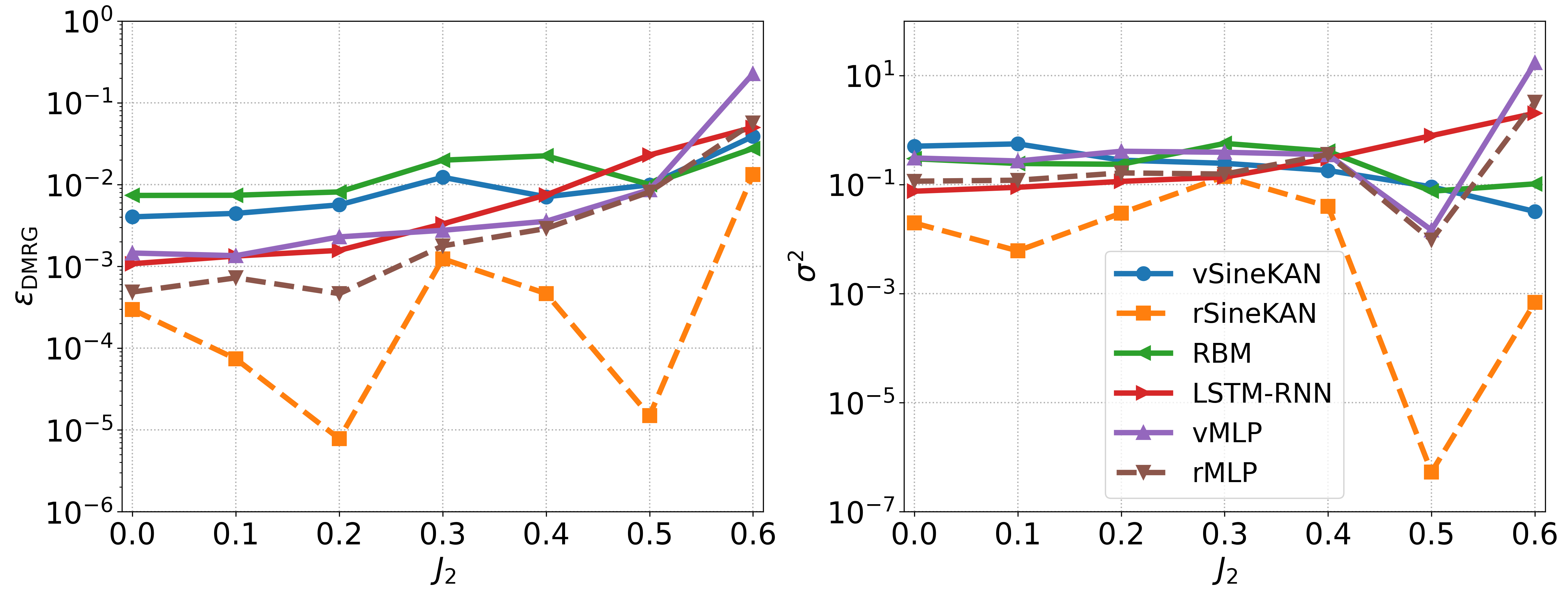}
    \caption{Comparison of error relative to DMRG $\epsilon_{\mathrm{DMRG}}$ (Left) and variance (Right) for \vs, \rs, \rbm, \lstm, \vm, and \rmlp for the chain with length $L=100$.}
    \label{fig:l100-compare}
\end{figure*}

We compute the relative error of the \sk GS energy with respect to the DMRG result for system sizes $L = 32, 64$ and $100$ as:

\begin{equation}\label{eq:1}
    \epsilon_{\mathrm{DMRG}} \equiv \frac{|E_{\mathrm{DMRG}}-E_{\mathrm{SineKAN}}|}{|E_{\mathrm{DMRG}}|} 
\end{equation}
and the variance $\sigma^2$ computed over a sample of configurations.

We compare the performance of \rs and \vs in terms of relative error (\cref{fig:relerror}) and variance (\cref{fig:variance}). We find that for smaller lattices of size $L=32$ and $L=64$, there is generally not a large substantial difference between \rs and \vs. However, for $L=100$, the addition of reflection symmetry in \rs leads to a significant improvement in relative error and variance. The approximate magnitude of relative error is one to three orders of magnitude less for \rs compared to \vs for $L=100$ and $J_2<0.6$ and the variance is approximately zero to five orders of magnitude less.

For \rs with size $L=100$, at $J_{2}=0.0, 0.4$ and $0.5$ the percentage errors ($100\times\epsilon_{\mathrm{DMRG}}$) are of the order of $\sim 0.03\%$, $\sim 0.04\%$ and $\sim 0.001\%$ respectively. The most significant improvement is at the MG point. At this point, the GS corresponds to exact VBS order. This means, since knowing one spin of each pair gives the other spin, the effective parameter space is reduced from $2^L$ to $2^{L/2}$. \rs is more able to effectively model this point.

We also directly compared, for $L=100$, the performance of the \rbm, \lstm, \vs, \rs, \vm, and \rmlp \textit{ansätze} for $J_2=0.0-0.6$. A comprehensive summary of findings of comparison of the \rs and \vs \textit{ansätze} to \rbm and \lstm can be found in \cref{fig:l100-compare}. Here we see that \rs outperforms the other \textit{ansätze} across the board on approximation of the DMRG GS energy and low variance with the sole exception of best variance for \lstm on $J_2=0.3$. We see the comparison of \rs and \vs to \vm and \rmlp in \cref{fig:l100-compare}. Here we see that \rs outperforms an \mlp with similar number of parameters. The inclusion of reflection symmetry leads to an improvement of the result for both \mlp and \sk but with the best overall result corresponding to \rs.

Previous work has used translationally invariant transformers to study the $L=100$ $J_1-J_2$ spin chain \cite{ViT1}. They found an approximate value of $\epsilon$ $\epsilon_{\mathrm{DMRG}}$ of $~1.65\times10^{-4}$ for $J_2=0.4$ and $~7.36\times10^{-6}$ for $J_2=0.0$. The errors for \rs were $4.64\times10^{-4}$ for $J_2=0.4$ and $2.96\times10^{-4}$ for $J_2=0.0$ respectively.

\section{MSR in SineKAN} \label{sec:msr}
It is known that the MSR holds exactly at $J_{2}=0$ and remain approximately valid for $0<J_{2} \leq 0.5$ in the 1D $J_1\text{--}J_2$ model; however, the sign structure of the GS becomes non-trivial for $J_2 > 0.5$ \cite{Beccarbm,Nikita}. There are three different ways of incorporating MSR into NQS architectures. One approach is to train the model without any prior encoding of MSR, allowing it to learn the sign structure over the course of optimization \cite{ViT1}. Another strategy involves using two separate NNs: one to represent the amplitude and another to learn the phase of the variational wave function, without explicitly enforcing MSR \cite{szabo}. A third method is to implement MSR at the Hamiltonian level together with a predefined, fixed phase structure embedded directly into the \textit{ansatz} \cite{CNN1}.

\begin{figure}[ht]
    \centering
    \includegraphics[width=\columnwidth]{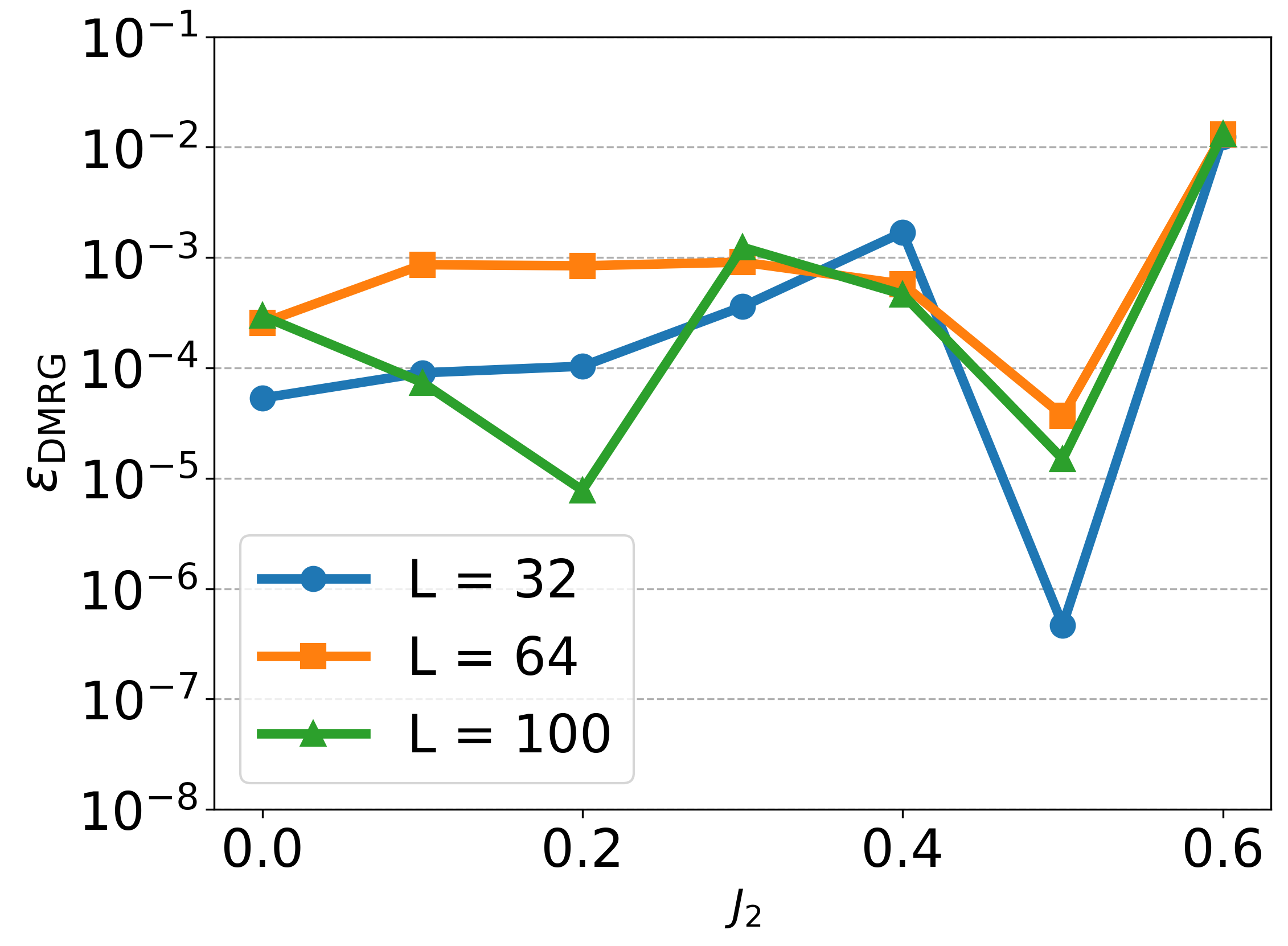}  
    \caption{\rs relative errors vs $J_2$ for system sizes $L=32, 64$, and $100$.}
    \label{fig:MSR}
\end{figure}

In our model, we incorporated the MSR in the Hamiltonian without introducing any additional phase structure in the variational wave function.  To determine how well our \textit{ansatz} captures the sign structure we plotted the $\epsilon_{\mathrm{DMRG}}$ as a function of $J_{2}$ for $L = 32, 64, 100$, as shown in (\cref{fig:MSR}). For $J_2 \leq 0.5$, the error $\epsilon_{\mathrm{DMRG}}$ remains consistently small across all system sizes, indicating that the MSR is well respected in this regime. This suggests that our \textit{ansatz} accurately captures the GS sign structure without requiring any additional phase learning.

In the highly frustrated regime at $J_{2}=0.6$ for the system sizes of $L = 32, 64, 100$, the GS energy computed by \rs has a percentage error approximately $1.3\%$ from the corresponding DMRG values, indicating that \rs struggles to accurately capture the sign structure in this region. For $L=20$, the fidelity between the \rs and the exact GS wavefunction decreases from $\sim100\%$ at $J_2<0.6$ to approximately $\sim 41\%$ at $J_2=0.6$. Additionally, the error ($\epsilon_{\mathrm{DMRG}}$) is roughly the same $\sim1.3\%$ for $L=32$, $64$, and $100$ as shown in \cref{fig:MSR}. The behavior of \rs is consistent for $J_2=0.6$ across different lattice sizes, and we attribute this directly to violation of MSR for $J_2>0.5$ and not directly to any particular choice of the \textit{ansatz}. This is supported by results shown in \cref{fig:l100-compare} and \cref{fig:MSR} where \rs has the best estimation of the DMRG energy result among \textit{ansätze} for $J_2=0.6$.

\section{Inference Time}\label{sec:inference}

\begin{figure}
    \centering
    \includegraphics[width=\linewidth]{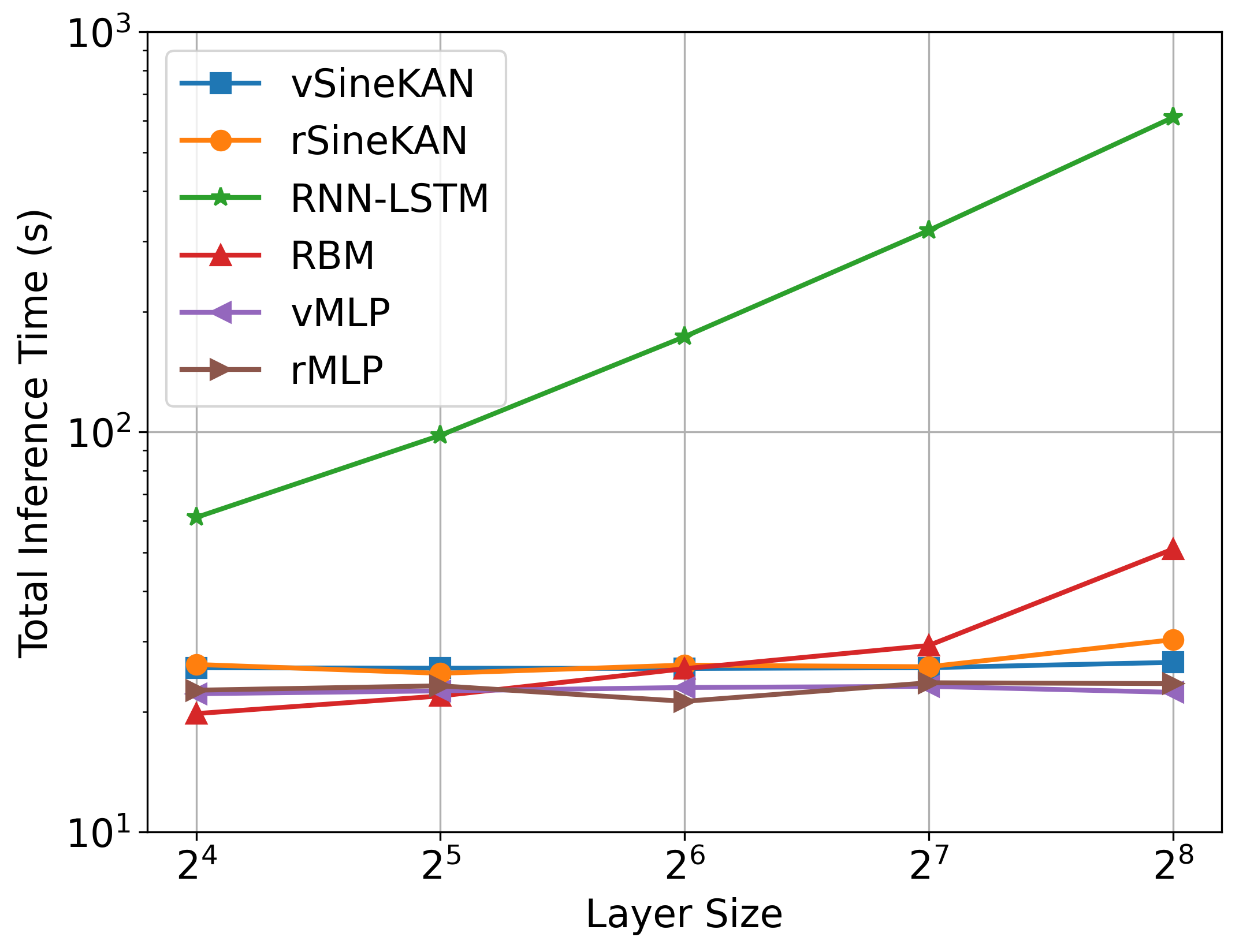}
    \caption{Inference time averaged over 100,000 forward steps for \vs, \rs, \lstm, \rbm, \vm, \rmlp. \sk and \mlp architectures show roughly flat scaling compared to linear scaling for \rbm and exponential scaling for the \rbm.}
    \label{fig:inftime}
\end{figure}

We compare inference times on a single forward pass with an input vector of size $(1, L)$ representing a single lattice of length $L$. We plot inference time as a function of $L$ for $L=16, 32, 64, 128, 256$. Results were averaged over 100,000 forward passes with each preceded by 200,000 warm up iterations to ensure consistent behavior on GPU. The GPU used was a single NVIDIA T4 GPU with 16GB of RAM. For the \rbm we use 128 features per node, for \lstm we use three hidden layers each with 64 features, for \vs and \rs we use a grid size of 8 and layer sizes of $(64,64,1)$, and for \vm and \rmlp we use layer sizes of $(256,256,1)$. We see in \cref{fig:inftime} which has $\log_{10}$ y-axis and $\log_2$ x-axis that \sk and \mlp architectures show roughly flat scaling and only deviate significantly at $L=256$ where the cost contribution from the first layer is large enough that linear scaling takes over. \lstm shows a linear growth in inference time and \rbm, while fastest for small lattice sizes, shows an exponential growth in inference time. \vm and \rmlp have faster inference than \vs and \rs.

\section{Discussion}\label{sec:discussion}

In this paper, we present the \sk model as a potential \textit{ansatz} for NQS wavefunctions. The motivation to explore KAN architectures stems from the KAST \cite{Kolmogorov1956, Kolmogorov1957} and supporting literature on the effectiveness of various KAN \textit{ansatz} \cite{comprehensiveKAN}.
The specific choice of \sk was inspired by its strong numerical performance in function approximation and scalability relative to other KAN variants \cite{sinekan}. Moreover, \sk has been shown to effectively generalize patterns beyond the training domain, further motivating its exploration \cite{sinekan}.

In \autoref{sec:results} we establish the efficacy of including reflection symmetry about the center of the lattice. Passing the input spins and the center-reflected input spins to the same multilayer \sk and then summing to make the combined \rs effectively halves the total parameter space and stabilizes performance at minimal cost. We establish, using multiple benchmarks of GS energy, correlation functions, and structure factors, that the \vs \textit{ansatz} is numerically accurate for modeling $L=20$ TFIM, AHM, and $J_1-J_2$ systems and that these results can be improved further by adding a minimal reflection symmetry in \rs. In particular, to resolve the near‐degeneracy of the two Néel product states for $\gamma > 0$ in the AHM, we introduced a small Zeeman bias $h_{\mathrm{init}} = \gamma + 0.2$ and annealed it to zero over the course of the VMC optimization. This bias lifts the twofold Néel degeneracy by an energy, selects a unique nondegenerate GS, and substantially improves convergence in the positive $\gamma-$regime. For the case AHM, \vs is unable to resolve the ground state and excited states from $\gamma=0.5$ to $0.9$ due to the vanishing energy gap. However, \rs is able to resolve the GS, likely due to the reflection symmetry reducing the total search space by a factor of two and stabilizing model behavior.

Furthermore, we show later in \autoref{sec:results} that for $J_1-J_2$ spin chain of $L=100$, including reflection symmetry leads to an improvement in performance for both \sk and \mlp-based architectures. It should be noted that the improvement in performance is more significant from \vs to \rs than for \vm to \rmlp. We also recommend that, for other lattice geometries, appropriate discrete symmetries be included. Given sufficient resources, this could even include computationally expensive symmetries such as translational invariance, which would require, as an upper limit, a number of forward passes equal to the lattice size.

In the \autoref{sec:results}, we show that the model achieves substantially more accurate results in estimation of the GS energy at lower variance across configurations when compared to benchmark \lstm, \mlp and \rbm models. We also find that \sk is highly computationally efficient compared to \lstm with similar numbers of parameters and numerically competitive with \mlp architectures with the same number of parameters and the same reflection symmetry. Based on these findings, we expect that \rs \textit{ ansatz} will be highly competitive with other \textit{ansätze} for NQS. A particular indicator of the efficacy of \rs is that it is the only \textit{ansatz} explored in this paper that is consistently able to model, with extremely low relative error and variance, the MG point in the $J_1-J_2$ model. This strongly motivates the choice of \rs as an \textit{ansatz} for NQS.

We also compare to previous results using translationally invariant vision transformers (ViT) \cite{ViT1}. It is difficult to directly compare these models due to differences in computational resources and algorithms used, however ViT outperforms \rs for $J_2=0.0$ and $0.4$. Due to the authors' reported significantly increased sample size and the use of the stochastic reconfiguration algorithm, we estimate training these models would have required, at minimum, on the order of twice the amount of compute resources in both RAM and time. Future work might explore the use of full translational invariance in combination with \sk layers. Furthermore, \sk layers could potentially be used in heads in transformer architectures, possibly in place of \mlp heads.

We incorporate the MSR into the $J_1$–$J_2$ Hamiltonian, which yields accurate results for \rs in the regime $J_2 = 0$ to $0.5$ for the chain length $L = 32, 64$, and $100$. However, the violation of the MSR for $J_2 = 0.6$, results in a noticeable drop in performance. In this regime, all other \textit{ansatz} also perform poorly, with the best result still coming from \rs, which maintains a consistent relative error with respect to DMRG, $\epsilon_{\mathrm{DMRG}} \sim 1.3\%$, independent of the chain length.

There exists an \textit{anomalous} case for \sk.
Due to symmetries in the phase shifts in \sk, when hidden layer dimension is sufficiently similar to $L$ and, simultaneously, the grid size is equal to $\sqrt{L}$, predictive performance of the models becomes substantially worse. This can be clearly seen in \cref{tab:failcase}. To account for this, we use a grid size of 7 rather than 8 for $L=64$. This is generally easily predictable and avoidable for cases where the layer sizes match the input dimension nearly exactly and simply require adjusting the grid size to be at least $\sqrt{N}\pm1$ and not exactly $\sqrt{N}$. We show that changing the grid size by one resolves this issue. We posit that this is likely related to specific alignments of the grid and input phase shifts but further exploration is needed to confirm this.

We also show the computational scaling for \sk architectures compared to other \textit{ansätze}. \sk has scaling much more comparable to the highly efficient \mlp architectures, while \lstm has larger polynomial scaling and \rbm has exponential scaling. Additionally, the \rs model is highly accurate even when using $1024$ samples in the configuration and sampling over 34K iterations. This represents only $\sim3.5\times10^7$ total spin configurations sampled from the total parameter space of $2^{100}\approx1.3\times10^{30}$. It's likely that increasing the configuration size and model parameters in future longer training runs on high-performance computing systems could lead to more accurate results.

Beyond the benchmark applications explored in this work, the \sk architecture holds promise for a wide range of quantum many-body systems. Its flexibility and expressivity make it a compelling candidate for studying critical phenomena in 1D quantum materials. Future work could include applications in ferromagnetic spin chains, magnetic polymer chains and aggregates \cite{random_ising, Korshak1987,aggregate}, and magnetic atomic chains—systems where entanglement features and nontrivial correlations play a pivotal role. Moreover, the model may offer valuable insights into avalanche-like magnetization dynamics in artificial spin ices, where nanomagnets are arranged on tailored lattices. The \rs \textit{ansatz} could also be extended to incorporate many-body interactions relevant to fermionic and bosonic systems, potentially extending its applicability to more complex condensed matter settings.

\section{ Conclusion \& Outlook}\label{sec:conclusion}
In this paper we introduce \sk as a potential \textit{ansatz} for the NQS approach for describing the GS of quantum many-body systems. We show strong empirical evidence that the model has a high degree of accuracy with low variance in computing GS energies and correlation functions of TFIM, AHM, and $J_1-J_2$ systems when compared to ED for a lattice size of $L=20$. We further show that, for lattice sizes of $L=32, 64$, and $100$, \sk models are highly accurate compared to the results from the DMRG algorithm. Changing only the input chain length, $L$, in the first layer of \sk, robust numerical results can be achieved with a linear increase in the model size as a function of $L$.

We benchmark \sk against \mlp, \rbm and \lstm models for a lattice size of $L=100$. The best performing model was found to be \rs which adds a reflection symmetry about the center with low computation cost increase. We find that the \rs model outperformed the \lstm \textit{ansatz} despite \lstm including both short and long distance relationships across the lattice and \rs using a fraction of the total computation cost. We also compare with \mlp architectures with a slightly larger number of parameters and found that \rs still significantly outperforms \rmlp both in terms of low error and low variance meaning that the improvement in performance cannot be solely attributed to including reflection symmetry. We also benchmark against \rbm which is a commonly used benchmark \textit{ansatz} for NQS due to physics motivations of \rbm having an Ising model structure. We find that \rs significantly outperforms a \rbm with a substantially larger number of parameters.

We show empirically that \sk is a promising \textit{ansatz} which is worth exploring for future work on NQS. The models trained in this work can all be trained in less than 24 hours on a single Nvidia T4 GPU and can be run offline for inference on a laptop using only a CPU and 16GB of RAM. We have strong reason to believe that significant improvements can be achieved using greater computational resources. Further exploration might include larger configuration sizes, larger models, and additional symmetries, as well as generalizations of these approaches to other lattice geometries.

A key motivation for exploring KAN-based NQS was to investigate whether incorporating the KAST structure leads to improved approximations of the GS energy in the highly frustrated regime. While our results show clear improvements over traditional architectures such as \rbm in the large $J_2$ region, the model can still become trapped in local minima. A possible future direction is to extend the \sk frequencies and amplitudes into the complex domain, a technique which has been shown in other works to lead to improvement in learning the sign structure of models in the highly frustrated regime \cite{ViT1,Beccarbm}.

\section{Code Availability}
The code used to generate the results for this paper can be found at \url{https://github.com/ereinha/SineKAN_NQS}.

\section*{Acknowledgments}
We thank Giuseppe Carleo, Kyoungchul Kong, Riccardo Rende, and Md Moshiur Rahman for their valuable comments and discussions. MAS and PTA are grateful to the National Science Foundation (NSF) for financial support under Grant No. [1848418]. This work was supported by the U.S. Department of Energy (DOE) under Award No. DE-SC0012447 (E.R. and S.G.). The Variational Monte Carlo and \sk architectures were implemented following the VarBench \cite{varbench} framework, using the NetKet \cite{vicentini2022netket} and JAX \cite{mpi4jax:2021} libraries. The DMRG code was implemented using the ITensor \cite{itensor} library.

\FloatBarrier
\bibliographystyle{apsrev4-2}
\bibliography{references}

\begin{table*}[ht]
\captionsetup{justification=raggedright,singlelinecheck=false}
\newcolumntype{P}[1]{>{\centering\arraybackslash}p{#1}}
\centering
\renewcommand{\arraystretch}{1.8}
\begin{tabular}{|c|P{1.5cm}|P{1.7cm}|P{1.2cm}|P{1.7cm}|P{1.7cm}|P{1.5cm}|P{1.2cm}|P{1.8cm}|P{1.8cm}|}
\hline
\multirow{2}{*}{\texttt{Chain Length}} & \multirow{2}{*}{$\mathbf{J_2}$} 
& \multicolumn{4}{c|}{\texttt{\rs}} 
& \multicolumn{4}{c|}{\texttt{\vs}} \\
\cline{3-10}
& &
\shortstack{\texttt{Hidden}\\\texttt{Layer}} & \shortstack{\texttt{Grid}\\\texttt{Size}} & \shortstack{\texttt{No. of}\\\texttt{Params}} & \shortstack{\texttt{Learning}\\\texttt{Rate}} 
& \shortstack{\texttt{Hidden}\\\texttt{Layer}} & \shortstack{\texttt{Grid}\\\texttt{Size}} & \shortstack{\texttt{No. of}\\\texttt{Params}} & \shortstack{\texttt{Learning}\\\texttt{Rate}} \\
\hline
 & 0.0 &  &  &  & $1 \times 10^{-3}$ &  &  &  & $1 \times 10^{-4}$ \\

 & 0.1 &  &  &  & $1 \times 10^{-3}$ &  &  &  & $1 \times 10^{-4}$ \\

 & 0.2 &  &  &  & $1 \times 10^{-3}$ &  &  &  & $1 \times 10^{-4}$ \\

32 & 0.3 & 64, 64 & 8 & $51100$ & $1 \times 10^{-4}$ & 64, 64 & 8 & $51100$ & $1 \times 10^{-4}$ \\

 & 0.4 &  &  &  & $1 \times 10^{-4}$ &  &  &  & $1 \times 10^{-4}$ \\

 & 0.5 &  &  &  & $1 \times 10^{-4}$ &  &  &  & $1 \times 10^{-4}$ \\

 & 0.6 &  &  &  & $1 \times 10^{-4}$ &  &  &  & $1 \times 10^{-5}$ \\ 
 \hline

 & 0.0 &  &  &  & $1 \times 10^{-3}$ &  &  &  & $1 \times 10^{-4}$ \\

 & 0.1 &  &  &  & $1 \times 10^{-3}$ &  &  &  & $1 \times 10^{-4}$ \\

 & 0.2 &  &  &  & $1 \times 10^{-3}$ &  &  &  & $1 \times 10^{-4}$ \\

64 & 0.3 & 64, 64 & 7 & $59300$ & $1 \times 10^{-3}$ & 64, 64 & 7 & $59300$ & $1 \times 10^{-4}$ \\

 & 0.4 &  &  &  & $1 \times 10^{-3}$ &  &  &  & $1 \times 10^{-4}$ \\

 & 0.5 &  &  &  & $1 \times 10^{-3}$ &  &  &  & $1 \times 10^{-4}$ \\

 & 0.6 &  &  &  & $1 \times 10^{-3}$ &  &  &  & $1 \times 10^{-5}$ \\
\hline
 & 0.0 &  &  &   & $1 \times 10^{-3}$ &  &  &   & $1 \times 10^{-4}$ \\

 & 0.1 &  &  &  & $1 \times 10^{-3}$ &  &  &  & $1 \times 10^{-4}$ \\

 & 0.2 &  &  &  & $1 \times 10^{-3}$ &  &  &  & $1 \times 10^{-4}$ \\

100 & 0.3 & 64, 64 & 8 & $86400$ & $1 \times 10^{-4}$ & 64, 64 & 8 & 86400 & $1 \times 10^{-5}$ \\

 & 0.4 &  &  &  & $1 \times 10^{-4}$ &  &  &  & $1 \times 10^{-5}$ \\

 & 0.5 &  &  &  & $1 \times 10^{-4}$ &  &  &  & $1 \times 10^{-5}$ \\

 & 0.6 &  &  &  & $1 \times 10^{-4}$ &  &  &  & $1 \times 10^{-5}$ \\

\hline
\end{tabular}
\caption{Hyperparameters for \vs and \rs architectures with the chain lengths $L = 32, 64$, and $100$ for different $J_2$ values.}
\label{tab:hyperparams}
\end{table*}

\FloatBarrier
\begin{appendix}

\section{Variational Monte Carlo}\label{sec:vmc}
Variational Monte Carlo (VMC) \cite{Sorella} is a powerful method for approximating the GS of a quantum system. It works in two steps. First, the quantum expectation value of the energy is cast into a classical Monte Carlo sampling problem over a probability distribution defined by a trial wavefunction that depends on a set of variational parameters. Second, the parameters of this wavefunction are optimized to minimize the estimated energy using stochastic optimization techniques.

For a trial wavefunction written in the computational basis as $\Psi_{\theta
} = \sum_{\sigma} \psi_{\theta}(\sigma) \ket{\sigma}$, which depends on a set of variational parameters $\theta$, the energy expectation value is given by 
\begin{equation}
E_{\theta} = \frac{\bra{\Psi_{\theta}}H\ket{\Psi_{\theta}}}{\braket{\Psi_{\theta}|\Psi_{\theta}}} \geq E_{0}.
\end{equation}
where $E_{0}$ the exact GS energy serving as the lower bound. Inserting a resolution of identity in the computational basis and defining $\psi_{\theta}(\sigma) = \braket{\sigma|\psi_{\theta}}$,  the energy expectation can be expressed as
\begin{equation}
    E_{\theta} = \sum_{\sigma} p_{\theta}(\sigma)  E_{\mathrm{loc}}(\sigma|\theta) = \langle\langle E_{\mathrm{loc}}(\sigma|\theta) \rangle\rangle,
\end{equation}
with $\langle\langle \cdots \rangle\rangle$ denoting the expectation value with respect to the $p_{\theta}(\sigma)$. Here the local energy is given by
\begin{equation}
    E_{\mathrm{loc}}(\sigma|\theta) = \sum_{\sigma'} \bra{\sigma'}H\ket{\sigma} \frac{\psi_{\theta}(\sigma')}{\psi_{\theta}(\sigma)},
\end{equation}
and is computed for each sampled configuration $\sigma$ and the probability distribution is defined as
\begin{equation}
p_{\theta}(\sigma) = |\psi_{\theta}(\sigma)|^{2}/\sum_{\sigma} |\psi_{\theta}(\sigma)|^{2}
\end{equation}
 For any observable, then, we can always compute expectation values over arbitrary wave functions as a statistical average. The great majority of physical observables, and for a given $\sigma$, the number of elements $\sigma'$ connected by those matrix elements is polynomial in the system size; thus, the summation in local energy can be carried out systematically. The essential idea of the VMC is, therefore, to replace these sums over exponentially many states, with a statistical average over a large but finite set of states sampled according to the probability distribution $p_{\theta}(\sigma)$.

Having introduced the Monte Carlo procedure for estimating the energy in large systems, the next step is to optimize the variational parameters. It is straightforward to show that the gradient of the energy can also be written under the form of the expectation value of some stochastic variable. In particular, define
\begin{equation}
D_{k}(\sigma) = \frac{1}{\Psi_{\theta}(\sigma)}\frac{\partial}{\partial\theta_{k}}\Psi_{\theta}(\sigma) = \frac{\partial}{\partial\theta_{k}}\log \Psi_{\theta}(\sigma)   
\end{equation}
the gradient of the energy expectation value with respect to the parameters $\boldsymbol{\theta}$ can
be written as
\begin{equation}
\frac{\partial}{\partial\theta_{k}} \langle\langle E_{\mathrm{loc}}(\sigma|\theta) \rangle\rangle = \langle\langle G_{k} \rangle\rangle
\end{equation}
with the statistical expectation value of the gradient estimator
\begin{equation}
    G_{k} = \langle\langle 2\mathrm{Re}[D_{k}(\sigma)-\langle\langle D_{k}(\sigma) \rangle\rangle]E_{\mathrm{loc}}(\sigma|\theta)
\end{equation}
The parameters can be iteratively updated using gradient descent until the energy converges.

In practice, the sampling of configurations $\sigma$ is typically performed using the Metropolis–Hastings algorithm, which generates a Markov chain of spin configurations distributed according to the probability $p_{\theta}(\sigma)$. Once a sufficiently large and equilibrated set of samples is collected, the gradient estimates can be computed. For the optimization step, simple stochastic gradient descent may be used, but more sophisticated techniques such as the stochastic reconfiguration (SR) \cite{Sorella,2024NatPh..20.1476C} method—also known as natural gradient descent—are often preferred. These take into account the underlying geometry of the variational manifold, leading to improved stability and convergence properties in high-dimensional parameter spaces.

\end{appendix}

\end{document}